\documentclass[aps,pra,twocolumn,superscriptaddress,groupedaddress]{revtex4-1}  % for review and submission

\usepackage{graphicx}  % needed for figures
\usepackage{dcolumn}   % needed for some tables
\usepackage{bm}        % for math
\usepackage{amssymb}   % for math
\usepackage{amsmath}
\usepackage{graphicx}
\usepackage{float}
\usepackage{hyperref}

\begin{document}

%%%%%%%%%%%%%%%%%%%%%%%%%%%%%%%%%%%%%%%%%%%%%%%%%%%%%%%%%%%%%%%%%%%%%%%
\title{Quantum Cascade Lasers: Electrothermal Simulation}
\author{S. Mei}
\author{Y. B. Shi}
\author{O. Jonasson}
\author{I. Knezevic}\email{iknezevic@wisc.edu}
\affiliation{Department of Electrical and Computer Engineering, University of Wisconsin-Madison, Madison,
WI 53706, USA}

\date{\today}

\begin{abstract}
Note: This is a book chapter that will appear in \textit{Handbook of Optoelectronic Device Modeling and Simulation}, Taylor \& Francis Books, 2017. Editor: Joachim Piprek. For table of contents, see \url{http://www.nusod.org/piprek/handbook.html}
\end{abstract}

\maketitle

%*****************************************************************************************
% Section 1, Introduction
\section{Introduction}\label{sec:intro}
Quantum cascade lasers (QCLs) are high-power, coherent light sources emitting in the mid-infrared (mid-IR) and terahertz (THZ) frequency ranges \cite{Faist94}. QCLs are electronically driven, unipolar devices whose active core consists of tens to hundreds of repetitions of a carefully designed stage. The QCL active core can be considered a superlattice (SL), in which each stage is a  multiple-quantum-well (MQW) heterostructure, where confined electronic states with specific energy levels are formed because of quantum confinement. The concept of achieving lasing in semiconductor SLs was first introduced by Kazarinov and Suris \cite{Kazarinov71} in 1971. The first working QCL was demonstrated by Faist \textit{et al.} \cite{Faist94} two decades later.

QCLs are typically III-V material systems grown on GaAs or InP substrates. Molecular beam epitaxy (MBE) \cite{Cheng97} and metal-organic chemical vapor deposition (MOCVD) \cite{Goetz83} are the techniques that enable precise growth of thin layers of various III-V alloys. It is also possible to incorporate strain into the structure, as long as the total strain in a stage is balanced. Both the precision and the possibility of introducing strain bring great flexibility to the design of the QCL active core, so lasing over a  wide range of wave lengths (from 3 to 190 $\mu$m) has been achieved. The growth techniques produce high-quality interfaces, with atomic-level roughness.

Mid-IR QCLs (wave-length range $3-12~\mu$m) have widespread military and commercial applications. A practical portable detector requires mid-IR QCLs to operate at room-temperature (RT), in continuous-wave (CW) mode, and with high (watt-level) output power. Furthermore, these QCLs must also have high wall-plug efficiency (WPE, the ratio of emitted optical power to the electrical power pumped in) and long-term reliability under these high-stress operating conditions. As the stress likely stems from excessive nonuniform heating while lasing \cite{Zhang10,Botez13},  improving device reliability and lifetime goes hand-in-hand with improving the WPE.

\subsection{Lasing in QCLs}\label{sec:lasing}
In QCLs, multiple conduction subbands are formed in the active core by means of quantum confinement. QCLs are unipolar devices, meaning that lasing is achieved through radiative intersubband transitions (transitions between two conduction subbands) instead of radiative interband transitions (transitions between the conduction and valence bands) in traditional quantum well (QW) semiconductor lasers. As a result, electrons do not combine with holes after the radiative transitions and can be used to emit another photon. In order to reuse electrons, the same MQW heterostructure is repeated many times (25--70) in the QCL active core (the so-called cascading scheme).

\begin{figure}
{\includegraphics[width=\columnwidth]{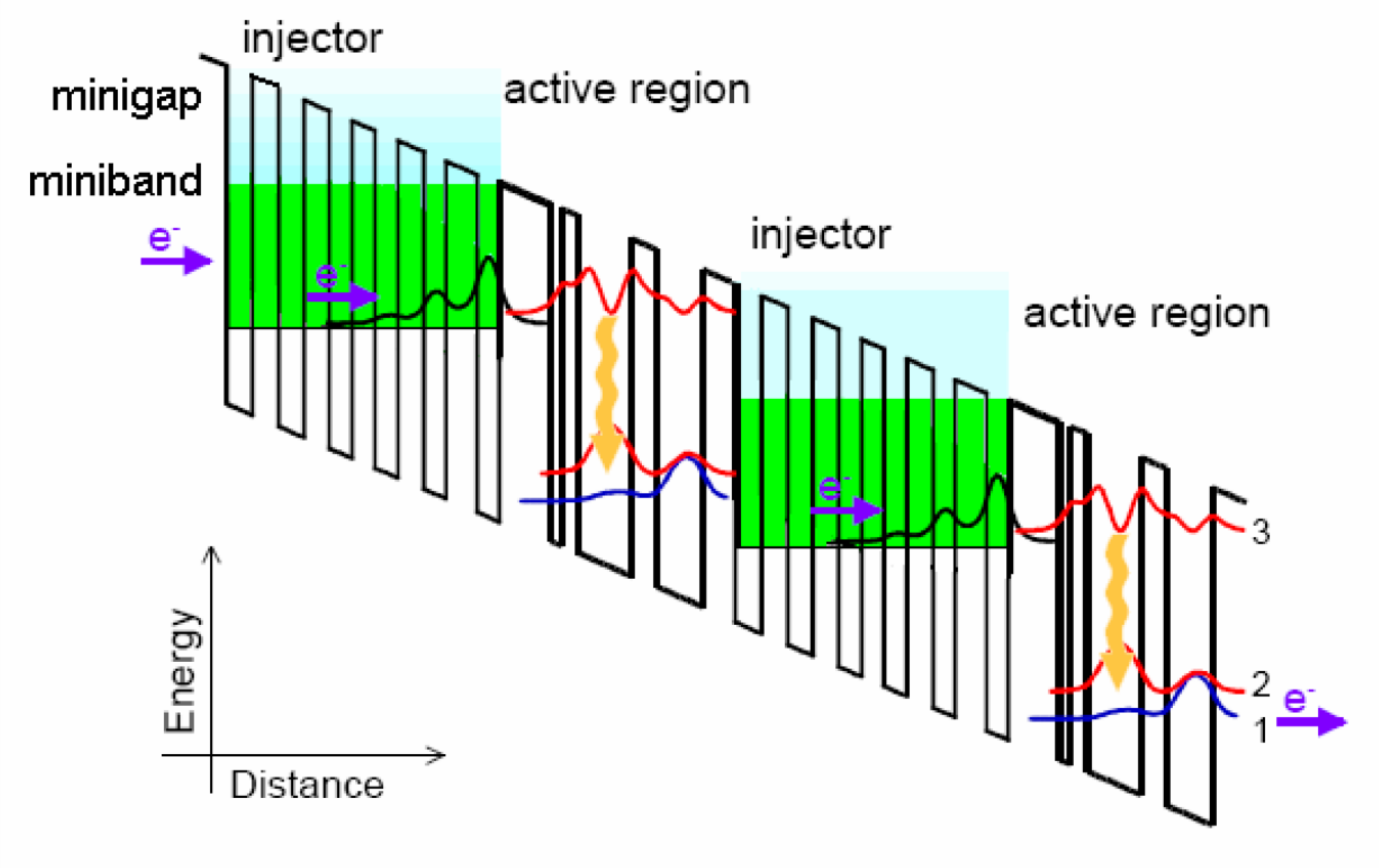}}
	\caption{A typical conduction-band diagram of two adjacent QCL stages under an applied electric field. Each stage consists of an injector region and an active region. A miniband is formed in the injector region while a minigap is formed in the active region (between the upper  and lower lasing levels). Lasing is associated with a radiative transition from the upper (3) to the lower  (2) lasing level. Electrons in the lower lasing level depopulate quickly to the ground level (1) by emission of longitudinal optical phonons.}\label{fig:bandstructure}
\end{figure}

Figure~\ref{fig:bandstructure} depicts a typical conduction-band diagram of two adjacent stages in a QCL under an electric field. Each stage consists of an injector region and an active region. The injector region has several thin wells separated by thin barriers ($10-30$ \AA), so a miniband is formed, with multiple subbands that are close in energy and whose associated wavefunctions have high spatial overlap. Typically, the lowest few energy levels in the miniband are referred to as the injector levels. The injector levels collect the electrons that come from the previous stage and inject them into the active region. The active region usually consists of 2--3 wider wells ($40-50$ \AA) separated by thin barriers. Consequently, a minigap forms in the active region between the upper lasing level (3) and the lower lasing level (2). Another important energy level in the active region is the ground state (1). There is a thin barrier (usually the thinnest among all layers) between the injecting region and the active region, called the injection barrier.

By design, the injector levels are close in energy and strongly coupled to the upper lasing level because of the thin injection barrier. The upper and lower lasing levels have large spatial overlap, which allows a radiative transition between the two levels; the wave length of the emitted light is determined by the energy spacing between these two levels. The lower lasing level overlaps with the ground state for efficient electron extraction. Electron emission of longitudinal optical (LO) phonons is the dominant mechanism for electron  extraction, so the energy spacing between the lower lasing level and the ground state is designed to be close to the LO phonon energy to facilitate extraction. With careful design, the electron lifetime in the upper lasing level is longer than in the lower lasing level, so population inversion can be achieved. After reaching the ground state, electrons tunnel through the injector into the the upper lasing level of the next stage, and the process is repeated. Of course, the lasing mechanism description above is idealized. In reality, the efficiency of the radiative transition between the upper and lower lasing levels is very low \cite{Liu10,Yao12}.

\subsection{Recent Developments in High-power QCLs}\label{sec:development}

In recent years, considerable focus has been placed on improving the WPE and output power of QCLs for RT CW operation. Bai \textit{et al.} \cite{Bai08} showed 8.4 \% WPE and 1.3 W output power around 4.6 $\mu$m in 2008. Shortly thereafter, Lyakh \textit{et al.} \cite{Lyakh09} reported 12.7\% WPE and 3 W power at 4.6 $\mu$m. Watt-level power with 6\% WPE at 3.76 $\mu$m and then lower power at 3.39 $\mu$m and 3.56 $\mu$m are reported by Bandyopadhyay \textit{et al.} \cite{Bandyopadhyay10,Bandyopadhyay12}. Bai \textit{et al.} \cite{Bai11} demonstrated 21\% WPE and 5.1 W output power around 4.9 $\mu$m in 2011. Much higher WPE and/or output power has been achieved at lower temperatures or at pulsed mode \cite{Liu10,Yao10} near 4.8 $\mu$m. A summary of recent developments can be found in review papers \cite{Yao12,Razeghi15}.

While good output powers and WPEs have been achieved, long-term reliability of these devices under RT CW operation remains a critical problem \cite{Zhang10,Botez13}. These devices are prone to catastrophic breakdown owing to reasons that are not entirely understood, but are likely related to thermal stress that stems from prolonged high-power operation \cite{Zhang10}. This kind of thermal stress is worst in short-wave length devices that have high strain and high thermal impedance mismatch between layers \cite{Wienold08,Lee10,Bandyopadhyay10,Bandyopadhyay12}.

In addition to improved device lifetime, we seek better CW temperature performance (higher characteristic temperatures $T_0$ and $T_1$, defined below) \cite{Botez13}. The first aspect is a weaker temperature dependence of the threshold current density. Empirically, the threshold current density (the current density at which the device starts lasing) has an exponential dependence on the operating temperature $T$: $J_\mathrm{th}\propto\exp\left(\frac{T}{T_0}\right)$. Higher characteristic temperature $T_0$ is preferred in QCL design, as it means less variation in $J_\mathrm{th}$ as the temperature changes.

Another key temperature-dependent parameter is the differential quantum efficiency (also called the slope efficiency of external quantum efficiency), defined as the amount of output optical power $dP$ per unit increase in the pumping current $dI$: $\eta_\mathrm{d}=\frac{dP}{dI}\propto\exp\left( -\frac{T}{T_1}\right)$. The differential quantum efficiency is directly proportional to the WPE ($\text{WPE}=\eta_d\eta_f$, where $\eta_f$ is the feeding efficiency). Therefore, the higher the $T_1$, the closer $\eta_\mathrm{d}$ is to unity, and the higher the WPE. Recently, deep-well structures with tapered active regions have demonstrated significant improvements in $T_0$ and $T_1$ with respect to the conventional $4.6\,\mu m$ device \cite{Bai08}, underscoring that the suppression of leakage plays a key role in temperature performance \cite{Kirch12,Botez13,Botez2016JPhysD}. Still, the microscopic mechanisms and leakage pathways that contribute to these empirical performance parameters remain unclear.

\subsection{QCL Modeling: An Overview}\label{sec:model}

Under high-power, RT CW operation, both electron and phonon systems in QCLs are far away from equilibrium. In such nonequilibrium conditions, both electronic and thermal transport modeling are important for understanding and improving QCL performance.

Electron transport in both mid-IR and THz QCLs has been successfully simulated via semiclassical (rate equations \cite{Indjin02,Indjin02a,Mircetic05} and Monte Carlo ~\cite{Iotti01,Callebaut04,Gao07,Shi14}) and quantum techniques (density matrix \cite{Willenberg03,kumar_PRB_2009,Weber09,Dupont10,terazzi_NJP_2010,Jonasson16,Jonasson16a}, nonequilibrium Green's functions (NEGF) \cite{Lee02,bugajski_PSS_2014,kolek_JSTQE_2015impact}, and lately Wigner functions \cite{jonasson15}). InP-based mid-IR QCLs have been addressed via semiclassical \cite{Matyas11} and quantum transport approaches (8.5-$\mu m$ \cite{lindskog_APL_2014} and 4.6-$\mu m$  \cite{bugajski_PSS_2014,kolek_JSTQE_2015impact} devices). There has been a debate whether electron transport in QCLs can be described using semiclassical models, in other words, how much of the current in QCLs is coherent. Theoretical work by Iotti and Rossi \cite{Iotti01,Iotti05} show that the steady-state transport in mid-IR QCLs is largely incoherent. Monte Carlo simulation \cite{Jirauschek09} has also been used to correctly predict transport near threshold. However, short-wavelength structures \cite{Bai08} have pronounced coherent features, which cannot be addressed semiclassically \cite{Jonasson16a}. NEGF simulations accurately and comprehensively capture quantum transport in these devices, but are computationally demanding. Density-matrix approaches have considerably lower computational overhead than NEGF, but are still capable of capturing coherent-transport features. A comprehensive review of electron-transport modeling was recently written by Jirauschek and Kubis \cite{Jirauschek14}.

 Electronic simulations that ignore radiative transitions are applicable for modeling QCLs below or near threshold, where the interaction between electrons and the laser electromagnetic field can be ignored. Such simulations are useful for predicting quantities such as threshold current density and $T_0$. However, in order to accurately model QCLs under lasing operations, the effect of the laser field on electronic transport would have to be included. In some cases, the effects of the laser field can be very strong~\cite{lindskog_APL_2014}, especially for high WPE devices, where the field-induced current can be dominant~\cite{Matyas11}. When included in simulations, the laser field is typically either modeled as an additional scattering mechanism~~\cite{jirauschek_APL_2010,Matyas11} or as a time-dependent sinusoidal electric field~\cite{wacker_JSTQE_2013,lindskog_APL_2014}. In this work, we ignore the effect of the laser field on electron dynamics.

Thermal transport in QCLs is often described through the heat diffusion equation, which requires accurate thermal conductivity in each region, a challenging task for the active core that contains many interfaces \cite{Evans08,Lee10,Lee12,Shi12,Mei15}. It is also very important to include nonequilibrium effects, such as the nonuniform heat-generation rate stemming from the nonuniform temperature distribution \cite{Shi12} and the feedback that the nonequilibrium phonon population has on electron transport \cite{Shi14}.

In this chapter, we present a multiphysics (coupled electronic and thermal transport) and multiscale (bridging between a single stage and device level) simulation framework that enables the description of QCL performance under far-from-equilibrium conditions \cite{Shi16_FdP}. We present the electronic (Sec.~\ref{sec:electron}) and thermal (Sec.~\ref{sec:thermal}) transport models,  then bring them together for electrothermal simulation of a real device structure (Sec.~\ref{sec:example}). We strive to cover the basic ideas while pointing readers to the relevant references for derivation and implementation details.

%*****************************************************************************************
% Section 2, Electronic transport

\section{Electronic Transport}\label{sec:electron}

Depending on the desired accuracy and computational burden, one can model electronic transport in QCLs with varying degrees of complexity. The goal is to determine the modal gain (proportional to the population inversion between the upper and lower lasing levels) under various pumping conditions (current or voltage) and lasing conditions (pulsed or continuous wave). A typical electron transport simulator relies on accurately calculated quasibound electronic states and associated energies in the direction of confinement. Electronic wavefunctions and energies are determined by solving the Schr\"{o}dinger equation or the Schr\"{o}dinger equation combined with Poisson's equation in highly doped systems. Section~\ref{sec:kp} introduces a $\mathbf{k\cdot p}$ Schr\"{o}dinger solver coupled with a Poisson solver. More information about other solvers for electronic states can be found in the review paper \cite{Jirauschek14} and references therein.

 The simulations of electronic transport fall into two camps depending on how the electron single-particle density matrix is treated. The diagonal elements of the density matrix  represent the occupation of the corresponding levels and off-diagonal elements represents the ``coherence'' between two levels. Transport is semiclassical or incoherent when the off-diagonal coherences are much smaller than the diagonal terms, and can be approximated as proportional to the diagonal terms times the transition rates between states \cite{Lee2006QCL}. In that case, the explicit calculation of the off-diagonal terms is avoided and only the diagonal elements are tracked, which simplifies the simulation considerably. However, when the off-diagonal terms are appreciable, transport is partially coherent and has to be addressed using quantum-transport techniques, discussed below.

\subsection{Semiclassical Techniques}\label{sec:semiclassical}
Semiclassical approaches assume that electronic transport between stages is largely incoherent ``hopping'' transport. The key quantities are populations of electronic states that are confined in the QCL growth direction, and electrons transfer between them due to scattering events. The scattering rates can be obtained empirically or more rigorously, via Fermi's golden rule. Common semiclassical approaches are the rate equations and ensemble Monte Carlo (EMC), the latter solving a Boltzmann-like transport equation stochastically.

\subsubsection{The Rate Equations}
In the rate-equation approach \cite{Indjin02,Indjin02a,Mircetic05}, scattering between relevant states, i.e., the injector level, the upper and lower lasing levels, and the ground state, is captured through transition rates. The rates include all relevant (radiative and nonradiative) scattering mechanisms, and can be either empirical parameters or calculated \cite{Faist02,Yamanishi08}. The computational requirements of rate-equation models are low, so they are suitable for fast numerical design and optimization of different structures \cite{Mircetic05}.

\subsubsection{Ensemble Monte Carlo}
The heterostructure in the QCL active core is a quasi-two-dimensional (quasi-2D) system, where electrons are free to move in the $x-y$ plane, while confined cross-plane, in the $z-$direction; the confinement results in the formation of quasibound states and discrete energy levels corresponding to the bottoms of 2D energy subbands. The electron wavefunctions in 3D are plane waves in the $x-y$ plane and confined wavefunctions in $z$. Electronic transport is captured by a Boltzmann-like semiclassical transport equation \cite{Iotti01}, which can be solved via the stochastic EMC technique assuming instantaneous hops between states in 3D due to scattering \cite{Suzeyth}. The simulation explicitly tracks the energy level and in-plane momentum of each particle in the simulation ensemble (typically $\sim 10^5$ particles). Tracking in-plane dynamics makes it more detailed than the rate-equation model. The transition rates are generally computed directly from the appropriate interaction Hamiltonians, and therefore depend on the energy levels as well as the wavefunction overlaps between different electronic states \cite{Suzeyth,Yanbingth,Jirauschek14}. EMC allows us to include nonequilibrium effect into transport, which is covered in more detail in Sec.~\ref{sec:EMC}.

\subsection{Quantum Techniques}\label{sec:quantum}
Density matrix and NEGF are the two most widely used techniques to describe quantum transport in QCLs. Recently, a Wigner-function approach was also successfully used to model a superlattice \cite{jonasson15}.

\subsubsection{Density-matrix approaches}\label{sec:DM}

In semiclassical approaches, the central quantity of interest is the distribution function $f_n^{E_k}(t)$, the probability of an electron occupying an eigenstate $n$ and having an in-plane kinetic energy $E_k$. The quantum-mechanical analogue is the single-electron density matrix, $\rho_{nm}^{E_k}(t)$, where the diagonal elements $\rho_{nn}^{E_k}(t)=f_n^{E_k}(t)$ are occupations and the off-diagonal elements $\rho_{nm}^{E_k}(t)$ are the spatial coherences between states $n$ and $m$ at the in-plane energy $E_k$. When employing semiclassical methods, off-diagonal matrix elements are assumed to be much smaller than diagonal elements. This approximation may fail in some cases, for example, when two eigenstates with a large spatial overlap have similar energies. This scenario often arises when modeling terahertz QCLs~\cite{callebaut_JAP_2005,kumar_PRB_2009,Jonasson16}, but can also come up in mid-IR QCLs~\cite{Jonasson16a}. In these cases, semiclassical models fail.

The density-matrix models that have been employed for QCL modeling can be categorized into two groups. The first includes hybrid methods, where transport is treated semiclassically within a region of the device (typically a single stage) while the effects of tunneling between different regions, separated by barriers, is treated quantum mechanically using a density-matrix formalism with phenomenological dephasing times~\cite{kumar_PRB_2009,terazzi_NJP_2010,callebaut_JAP_2005}. The second group involves completely quantum-mechanical methods that rely on microscopically derived Markovian master equations that guarantee positivity of the density matrix~\cite{Jonasson16,Jonasson16a}. Both methods are more computationally expensive than their semiclassical counterparts, because the density matrix contains many more elements than its diagonal semiclassical analogue.

\subsubsection{Nonequilibrium Green's functions}\label{sec:NEGF}

The nonequilibrium Green's function technique (see a good overview in \cite{Jirauschek14}) relies on the relationships between single-particle time-ordered Green's functions and correlation functions \cite{Lee02,bugajski_PSS_2014,kolek_JSTQE_2015impact}. The correlation function $G_{\alpha,\beta}^<(k;t_1,t_2)$, often referred to as the lesser Green's function ~\cite{Wacker02,wacker_JSTQE_2013}, is one of the central quantities and can be understood as a two-time generalization of the density matrix, where $k$ refers to the magnitude of in-plane wave vector. The correlation function contains both spatial correlations (terms with $\alpha \neq \beta$) as well as temporal correlations between times $t_1$ and $t_2$ (not included in semiclassical or density-matrix models). Typically, the potential profile is assumed to be time independent, in which case the correlation function only depends on the time difference $G_{\alpha,\beta}^<(k;t_1,t_2)=G_{\alpha,\beta}^<(k;t_1-t_2)$. Fourier transforming over the time difference into the energy domain gives the energy-resolved correlation function $G_{\alpha,\beta}^<(k,E)$, which is the quantity which is usually solved for numerically~\cite{Wacker02,wacker_JSTQE_2013,bugajski_PSS_2014}. The main advantages of the NEGF formalism are that it provides spectral (energy-resolved) information and it includes the effects of collisional broadening (the broadening of energy levels due to scattering), which is particularly important when the states are close in energy. These advantages carry a considerable computational cost, so NEGF calculations are much more time consuming than density-matrix approaches~\cite{lindskog_APL_2014}.

\subsection{Ensemble Monte Carlo with Nonequilibrium Phonons}\label{sec:EMC}

Here, we focus on presenting semiclassical modeling of electron transport in QCL structures via EMC \cite{Gao06,Gao07,Shi14}. The solver consists of two parts, a coupled Schr\"{o}dinger--Poisson solver and a transport kernel. We solve for the electronic states using the coupled Schr\"{o}dinger--Poisson solver and feed the energy levels and the wavefunctions of the relevant electronic states to the transport kernel. The transport kernel keeps track of the electron momentum, energy, and distribution among subbands. If the electron density inside the device is high, transport kernel will periodically feed the electron distribution back to the Schr\"{o}dinger--Poisson solver and update the electronic states. This loop is repeated until the electron distribution converges. By doing so, we solve for both electron transport and the electronic band structure self-consistently.

Since the active QCL core consists of repeated stages, the  wavefunctions in any stage can be obtained from the wavefunctions in any other stage by translation in space and energy. This translational symmetry makes it possible to simulate electron transport in only one generic central stage instead of in the whole QCL core \cite{Suzeyth}. Typically, electronic states in nonadjacent stages have negligible overlap, which also means that the transition rates between them are negligible. As a result, it is sufficient to limit interstage scattering events to only those between adjacent  stages.

\begin{figure}
	{\includegraphics[width=\columnwidth]{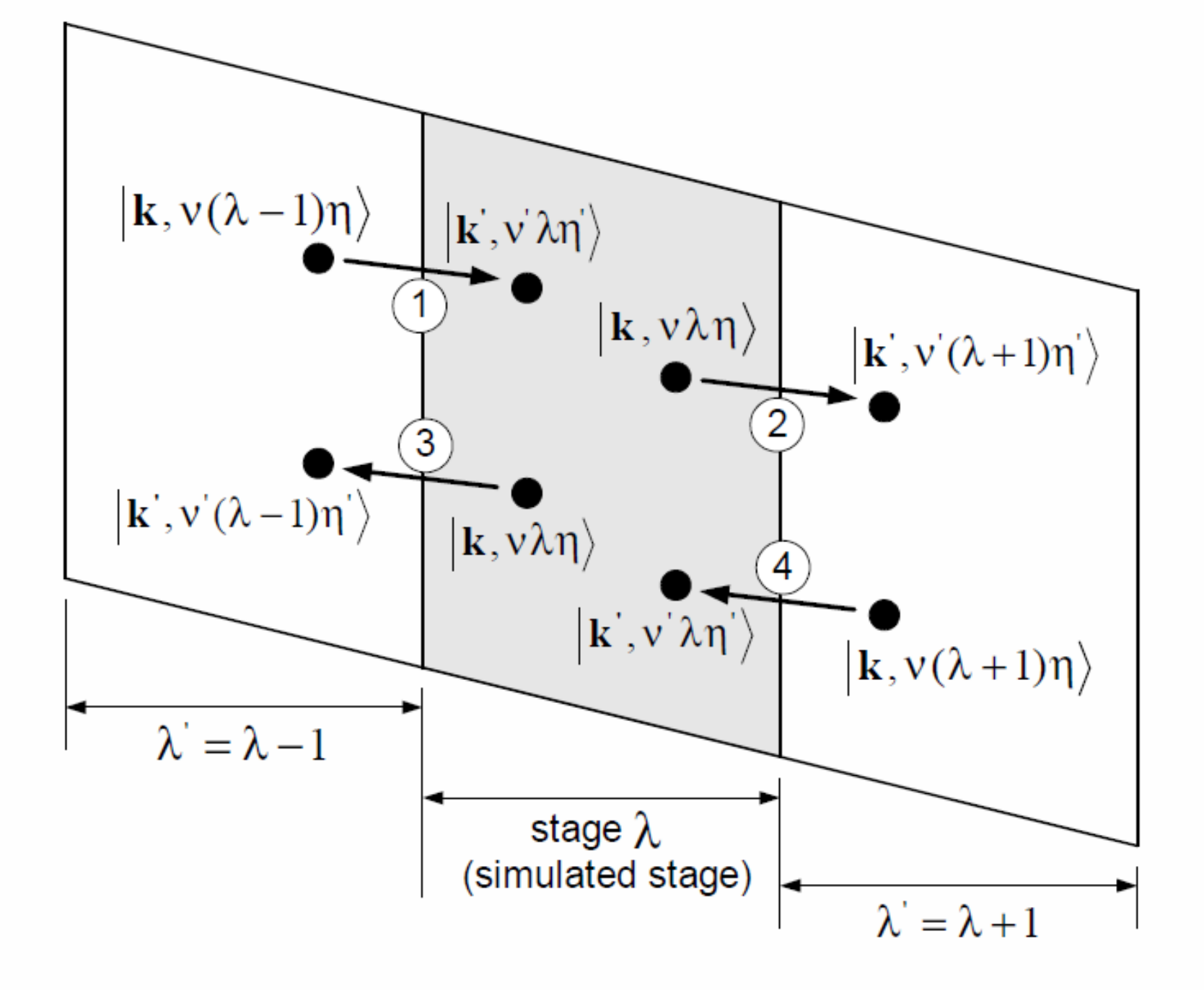}}
	\caption{Schematic of the three simulated stages in a QCL active core under an applied field. Scattering is limited to nearest-neighbor stages and periodic boundary conditions are justified by the cascading scheme; therefore, only three stages are needed in the EMC transport kernel.}\label{fig:EMCsche}
\end{figure}

Figure~\ref{fig:EMCsche} shows a schematic of three adjacent stages under an applied field. We simulate electron transport in the central stage $\lambda$, while nearest-neighbor interstage ($\lambda \rightleftarrows\lambda\pm 1$) and intrastage ($\lambda\rightarrow\lambda$) scattering is allowed. Periodic boundary conditions (PBCs) are applied in the simulation, i.e., whenever one electron scatters from the central stage out to the next stage (process $\textcircled{2}$), an electron scatters from the previous stage into the central stage (process $\textcircled{1}$) and vice versa (process $\textcircled{3}$ and process $\textcircled{4}$). PBCs are justified by the cascading scheme.

\subsubsection{Electronic Bandstructure Solver}\label{sec:kp}

We employ the $\mathbf{k\cdot p}$ method to solve the Schr\"{o}dinger equation and couple it to a Poisson solver \cite{Gao06,Gao07,Suzeyth}. The $\mathbf{k\cdot p}$ method is an efficient way to solve for the electronic band structure near the band edges, where the transport happens in QCLs. The $\mathbf{k\cdot p}$ method considers the contribution from the conduction band (C), light-hole band (LH), and the spin-orbit split-off band (SO) (the heavy-hole band (HH) decouples from the other three at the band edge) \cite{Chuang95}. The contributions from the LH and SO are especially important for narrow-gap materials, such as InP. Moreover, in modern QCLs, strain-balanced structures have been employed to obtain enhanced performance. In these structures, alternate layers are compressively or tensilely strained while the whole structure is strain free, with carefully designed thickness of each layer. The $\mathbf{k\cdot p}$ method allows for convenient inclusion of the effects of strain on the band structure. The implementation details of the $\mathbf{k\cdot p}$ solver can be found in \cite{Suzeyth}.

The $\mathbf{k\cdot p}$ solver can only solve for a finite structure rather than an infinite periodic one. As a result, we need to simulate a finite number of stages and add artificially high barriers to the two ends to confine all the states. If a stage is far enough from the boundaries, the calculated band structure in it should be the same as if we were to solve for the whole periodic structure. Tests have confirmed that three stages, which we also use in EMC, are enough when solving for the electronic states to ensure that the central-stage states are unaffected by the simulation-domain potential boundaries. The states from the central stage are then translated in energy and position to the neighboring stages according to the stage length and the applied electric field.

When we need to solve for electron transport and electronic states self-consistently, it is necessary for the solver to be able to automatically pick out the electronic states belonging to the central stage. One intuitive criterion is to calculate the ``center of mass'' for each state (the expectation value of the cross-plane coordinate, $\langle z\rangle$) and assign those falling in the central stage to that stage. However, in our three-stage scheme, this method may pick up the states that are too close to the boundary. One can either extend the number of stages in the $\mathbf{k\cdot p}$ solver to five, so the three stages in the middle are all far from the boundary, or use additional criteria, such as that there be more than 50\% possibility of finding an electron in the central stage, based on the probability density distribution, or requiring that the location of the probability-density peak be in the central stage. Additional criteria requiring strong confinement of states have been explored in \cite{Jonasson16}.

\subsubsection{Transport Kernel with Nonequilibrium Phonons}\label{sec:kernel}
The EMC kernel tracks the hopping transitions of electrons between subbands and stages until convergence, and outputs the transport information for us to calculate the experimentally relevant quantities such as current and modal gain \cite{Suzeyth}. In the transport kernel, both electron--electron interactions and electron--LO-phonon interactions are considered. Other scattering processes such as intervalley scattering, impurity scattering, interface roughness scattering can be considered under different circumstances \cite{Jirauschek14}. Photon emission is not considered, either.   Because EMC tracks individual particles, nonequilibrium electron transport can be automatically captured. (EMC tracks individual simulation particles, each of which might represent thousands of real electrons.)

The most important scattering mechanism in QCLs is electron--LO-phonon scattering, which facilitates the depopulation of the lower lasing level. As shown in \cite{Gao08}, phonon confinement has little effect on the electronic transport, therefore, for simplicity, LO phonons are treated as bulklike dispersionless phonons with energy $\hbar\omega_0$. The transition rate between an initial state $\phi_i(z)$ with energy $E_i$ and a final state $\phi_f(z)$ with $E_f$ can be derived from Fermi's golden rule as
\begin{eqnarray}
&&\Gamma_\mathrm{a(-),e(+)}=\frac{e^2\hbar\omega_0m_f^*}{8\pi^2\hbar^3}\left(\frac{1}{\epsilon_\infty}-\frac{1}{\epsilon_0}\right)\times\\
&&\int_{0}^{2\pi}d\theta\int_{-\infty}^{\infty}dq_{z}\int_{0}^{\infty}dE_{kf}
N_{\mathbf{q}}\frac{|\mathcal{I}_\mathrm{if}(q_z)|^2}{\mathbf{q}_\parallel^2+q_z^2}\delta(E_f-E_i\mp\hbar\omega_0),\nonumber
\end{eqnarray}
where $e$ is the electronic charge while $\epsilon_0$ and $\epsilon_\infty$ are static and high-frequency electronic permittivities of the material, respectively. The integrals are over the in-plane kinetic energy $E_k$ of the final state and the cross-plane momentum transfer $q_z$. $\mathbf{q_\parallel}=\mathbf{k'_\parallel}-\mathbf{k_\parallel}$ is the in-plane momentum transfer.
\begin{equation}\label{eq:OI}
|\mathcal{I}_\mathrm{if}(q_z)|^2=\left|\int_{0}^{d}dz\phi_f^*(z)\phi_i(z)e^{-izq_z}\right|^2
\end{equation}
is defined as the overlap integral (OI) between the initial and final states, where $q_z$ is the cross-plane momentum transfer. The integration is over the angle between initial and final in-plane momenta $\mathbf{k_\parallel}$ and $\mathbf{k'_\parallel}$ ($\theta$), cross-plane momentum component of the final state ($k_{fz}$), and the kinetic energy of the final state ($E_{kf}$). $N_\mathbf{q}$ represents the number of LO phonons with momentum $\mathbf{q}=(\mathbf{q_\parallel},q_z)$. The expression can be further simplified in the equilibrium case, where $N_\mathbf{q}$ follows the Bose-Einstein distribution \cite{Suzeyth}. In order to model nonequilibrium phonon effects, we numerically integrate the expression using a phonon number histogram according to both $q_\parallel$ and $q_z$ \cite{Yanbingth}.

\begin{figure}
	{\includegraphics[width=\columnwidth]{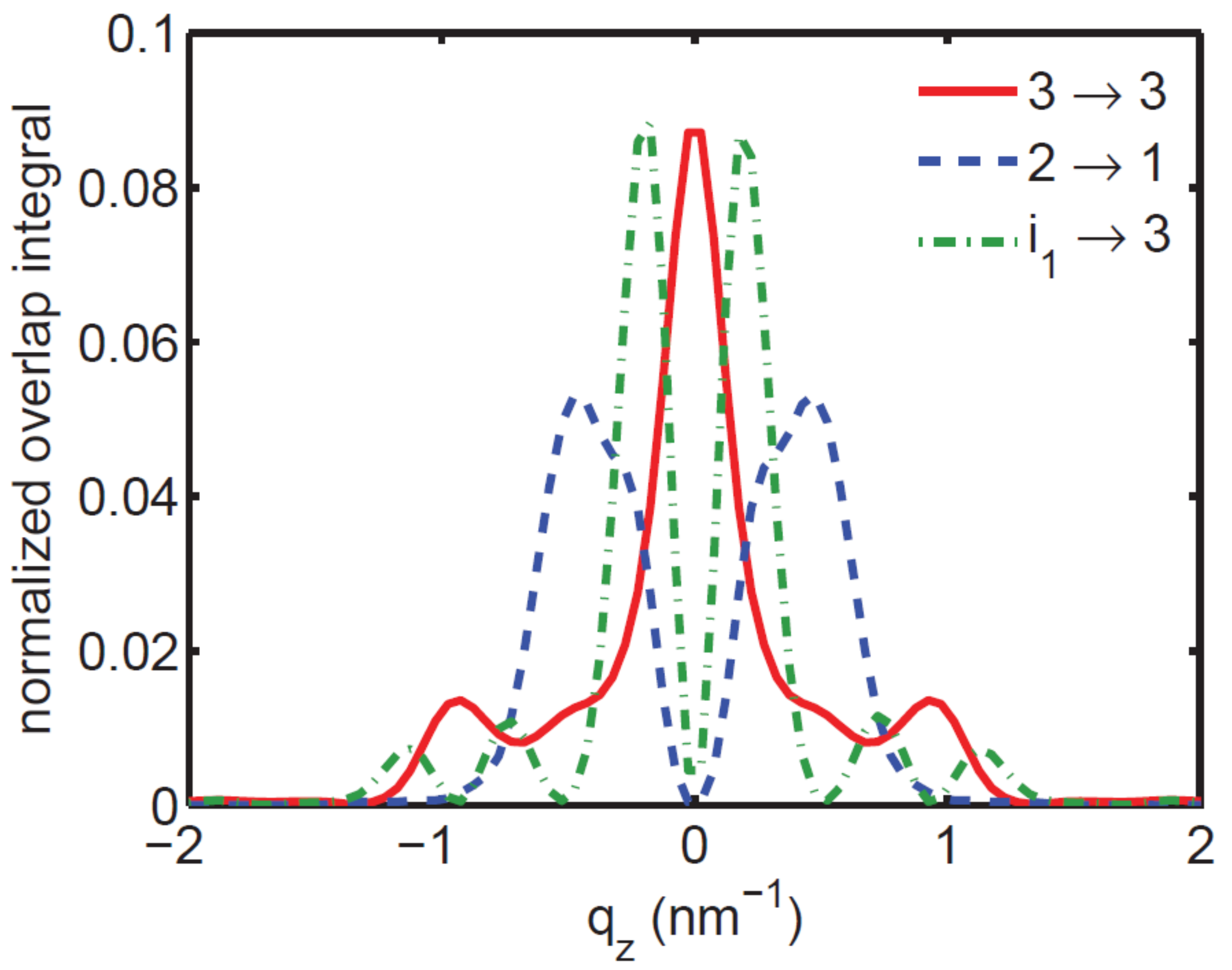}}
	\caption{Normalized overlap integral $|\mathcal{I}_\mathrm{if}|^2$ from Eq. (\ref{eq:OI}) versus cross-plane phonon wave vector $q_z$ for several transitions (intersubband $i_1\rightarrow 3$ and $2\rightarrow1$; intrasubband $3\rightarrow3$). Figure reproduced from \cite{Shi14}, Y. B. Shi and I. Knezevic, J. Appl. Phys. 116, 123105 (2014), with the permission of AIP Publishing.}\label{fig:overlap}
\end{figure}

According to the uncertainty principle, position and momentum cannot both be determined simultaneously. Since our electrons are all confined in the central stage ($\Delta z'$ is finite), the cross-plane momentum is not exactly conserved during the scattering process ($q_z\neq k'_z-k_z$) \cite{Lugli89}. This analysis does not affect the momentum conservation in the $x-y$ plane, because we assume infinite uncertainty in position there. Previously, the cross-plane momentum conservation has been considered through the momentum-conservation approximation (MCA) \cite{Ridley82,Lv06} and a broadening of $q_z$ according to the well width \cite{Lugli89}. The MCA forbids a phonon emitted between subbands $i$ and $f$ to be re-absorbed by another transition between $i'$ and $f'$ if $i\neq i'$ or $f\neq f'$, and thus might underestimate the electron-LO interaction strength \cite{Yanbingth}. The concept of well width is hard to apply in a MQW structure such as the QCL active core \cite{Yanbingth}. We observe that the probability of a phonon with cross-plane momentum $q_z$ being involved in an interaction is proportional to the overlap integral in Equation~(\ref{eq:OI}). Figure~\ref{fig:overlap} depicts the typical overlap integrals for both intersubband ($i_1\rightarrow 3$ and $2\rightarrow1$) and intrasubband ($3\rightarrow3$) transitions. As a result, in each electron--LO-phonon scattering event, we randomly select a $q_z$ following the distribution from the overlap integral (Fig.~\ref{fig:overlap}). Depending on the mechanism (absorption or emission), a phonon with ($\mathbf{q_\parallel},q_z$) is removed/added to the histogram according to the 2D density of states (DOS) and the effective simulation area~\cite{Yanbingth}. Once the phonons with a certain momentum are depleted, transitions involving such phonons become forbidden.

In order to couple the EMC solver to the thermal transport solver, we need to keep a detailed log of heat generation during electron transport. In all the relevant scattering events, electron--LO-phonon scattering is the only inelastic mechanism and therefore is the only mechanism that contributes to heat generation. As a result, the total energy emitted and absorbed in the form of LO phonons is recorded during each step of the EMC simulation. The nonequilibrium phonons decay into acoustic longitudinal acoustic (LA) phonons via a three-phonon anharmonic decay process. The formulation and the parameters here follow \cite{Usher94}. The simulation results of EMC including nonequilibrium phonons are shown in Section~\ref{sec:example}.

%*****************************************************************************************
% Section 3, Thermal Transport
\section{Thermal Transport}\label{sec:thermal}
\begin{figure}
	{\includegraphics[width=4cm]{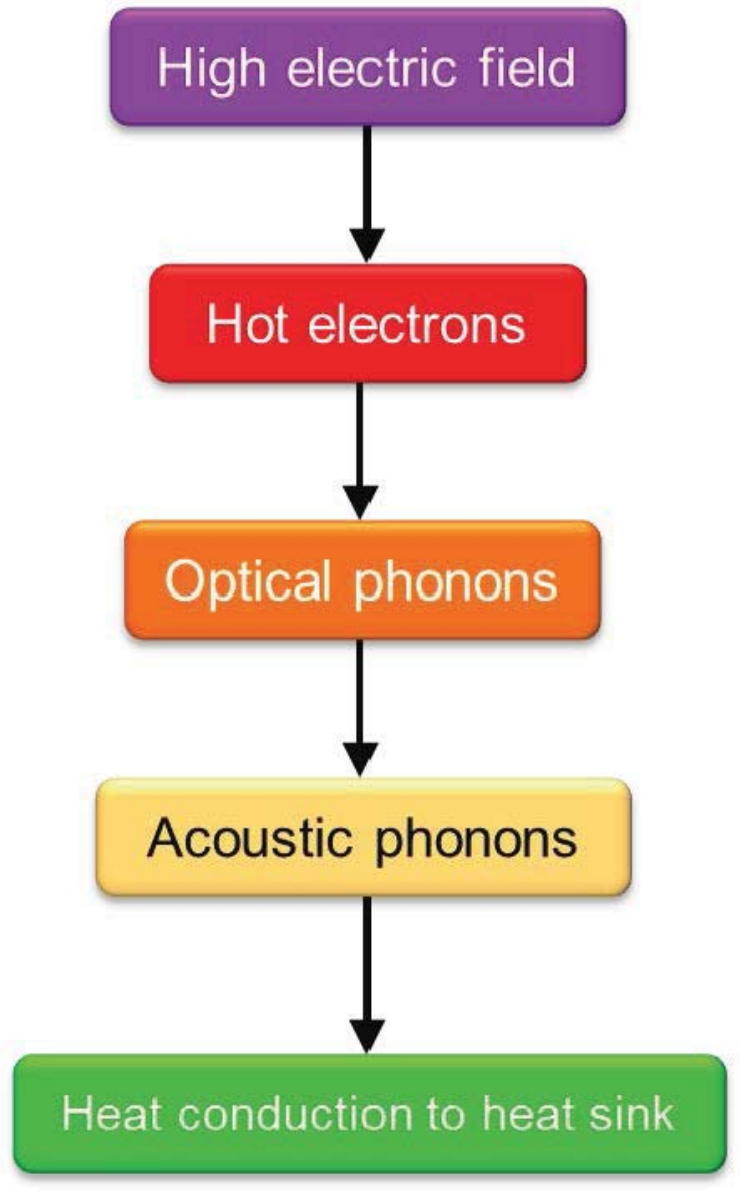}}
		\caption{Flow of energy in a quantum cascade laser.}\label{fig:heatflow}
\end{figure}

The dominant path of heat transfer in a QCL structure is depicted in Fig.~\ref{fig:heatflow}. The operating electric field of a typical QCL is high, which means that considerable energy is pumped into the electronic system.  These energetic, ``hot'' electrons relax their energy largely by emitting LO phonons. LO phonons have high energies but flat dispersions, so their group velocities are low and they are poor carriers of heat. An LO phonon decays into two LA phonons via a three-phonon process referred to as anharmonic decay. LA phonons have low energy but high group velocity and are the main carriers of heat in semiconductors \cite{Usher94,Shi12}. If we neglect the diffusion of optical phonons, the flow of energy in a QCL can be described by the equations
\begin{subequations}
\begin{eqnarray}
\frac{\partial W_A}{\partial t}&=&\nabla\cdot(\kappa_A\nabla T_A)+\left.\frac{\partial W_{LO}}{\partial t}\right|_\mathrm{coll}\, ;\\
\frac{\partial W_{LO}}{\partial t}&=&\nabla\cdot(\kappa_A\nabla T_A)+\left.\frac{\partial W_e}{\partial t}\right|_\mathrm{coll}-\left.\frac{\partial W_{LO}}{\partial t}\right|_\mathrm{coll}\, ,
\end{eqnarray}
\end{subequations}
where $W_{LO}$, $W_A$, and $W_e$ are the LO phonon, acoustic phonon, and electron energy densities, respectively. $\kappa_A$ is the thermal conductivity in the system and $T_A$ is the acoustic-phonon (lattice) temperature. The term $\nabla\cdot(\kappa_A\nabla T_A)$ describes heat diffusion, governed by acoustic phonons. We have also used the fact that the rate of increase in the LO-phonon energy density equals the difference between the rate of its generation by electron--LO-phonon scattering and the rate of anharmonic decay into LA phonons.

In a nonequilibrium steady state, both the LO and LA energy densities are constant, so
\begin{align}\label{eq:heatdiff}
-\nabla\cdot(\kappa_A\nabla T_A)=\left.\frac{\partial W_e}{\partial t}\right|_\mathrm{coll}.
\end{align}
As described in the previous section, the right-hand side of Equation~(\ref{eq:heatdiff}) is the heat-generation rate $Q$ and can be obtained by recording electron--LO-phonon scattering events in electronic EMC \cite{Pop06,Shi12}
\begin{equation}\label{eq:heatgen}
Q=\left.\frac{\partial W_e}{\partial t}\right|_\mathrm{coll}=\frac{N_\mathrm{3D}}{N_\mathrm{sim}t_\mathrm{sim}}\sum(\hbar\omega_\mathrm{ems}-\hbar\omega_\mathrm{abs}),
\end{equation}
where $N_\mathrm{3D}=\frac{N_s}{D_\mathrm{stage}}$ is the electron density ($N_s$ is the sheet density and $D_\mathrm{stage}$ is the length of a single stage) while $N_\mathrm{sim}$ and $t_\mathrm{sim}$ are the number of simulation particles and the simulation time, respectively.  $\hbar\omega_\mathrm{ems}$ and $\hbar\omega_\mathrm{abs}$ are the energies of the emitted and absorbed LO phonons, respectively. To solve Equation~(\ref{eq:heatdiff}), we need information on both the thermal conductivity $\kappa_A$ and the heat-generation rate $Q$; they are discussed in Subsections~\ref{sec:IIIVSL} and \ref{sec:device}, respectively.

%%%%%%%%%%%%%%%%%%%%%%%%%%%%%%%%%%%%%%%
\subsection{Thermal Conductivity in a QCL Device}\label{sec:IIIVSL}
%%%%%%%%%%%%%%%%%%%%%%%%%%%%%%%%%%%%%%%
\subsubsection{Active Core: A III-V Superlattice}
%%%%%%%%%%%%%%%%%%%%%%%%%%%%%%%%%%%%%%%

The QCL active core is a SL: it contains many identical stages, each with several thin layers made from different materials and separated by heterointerfaces. The thermal-conductivity tensor of a SL system reduces to two values: the in-plane thermal conductivity $\kappa_\parallel$ (in-plane heat flow is assumed isotropic) and the cross-plane thermal conductivity $\kappa_\perp$. Experimental results have shown that, in SLs, the thermal conductivity is very anisotropic \cite{Cahill14} ($\kappa_\parallel\gg\kappa_\perp$) while both $\kappa_\parallel\text{ and }\kappa_\perp$ are smaller than the weighted average of the constituent bulk materials \cite{Yao87,Chen94,Yu95,Capinski96,Capinski99}. Both effects can be attributed to the interfaces between adjacent layers \cite{Chen97,Chen98}.

Here, we discuss a semiclassical model for describing the thermal-conductivity tensor of III-V SL structures. Note that the model described here is in principle applicable to SLs in other material systems, as long as they have high-quality interface and thermal transport is mostly incoherent \cite{Huxtable02,Wang14_PRB,Wang15_APL,Mei15}. In particular, we focus on thermal transport in III-arsenide-based SLs, as they are most commonly used in mid-IR-QCL active cores~\cite{Mei15}.

Under QCL operation conditions of interest ($>77$ K, and typically near RT), thermal transport is dominated by acoustic phonons and is governed by the Boltzmann transport equation (BTE). To obtain the thermal conductivity, we solve the phonon BTE with full phonon dispersion in the relaxation-time approximation~\cite{Mei15}.

\subsubsection{Twofold Influence of Effective Interface Roughness}
\begin{figure}
	{\includegraphics[width=\columnwidth]{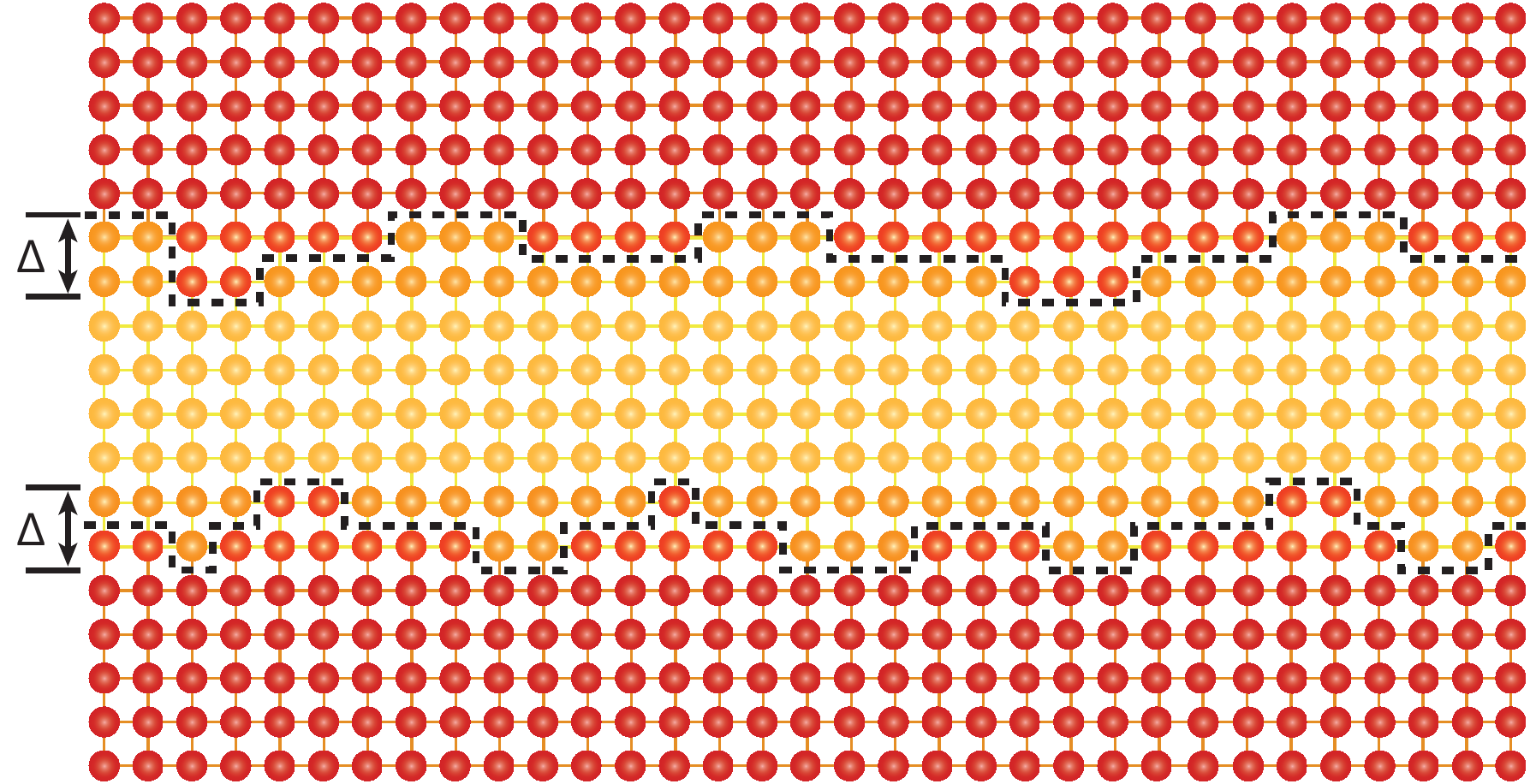}}
	\caption{Even between lattice-matched crystalline materials, there exist nonuniform transition layers that behave as an effective atomic-scale interface roughness with some rms roughness $\Delta$. This effective interface roughness leads to phonon-momentum randomization and to interface resistance in cross-plane transport. Figure reproduced from \cite{Mei15}, S. Mei and I. Knezevic, J. Appl. Phys. 118, 175101 (2015), with the permission of AIP Publishing.}\label{fig:SLsche}
\end{figure}

To capture both the anisotropic thermal transport and the reduced thermal conductivity in SL systems, we need to observe the twofold influence of the interface. First, it reduces $\kappa_\parallel$ by affecting the acoustic-phonon population close to the interfaces \cite{Aksamija13}. Second, it introduces an interface thermal boundary resistance (ITBR), which is still very difficult to model \cite{Swartz89,Cahill14}. Common models are the acoustic mismatch model (AMM) and the diffuse mismatch model (DMM)~\cite{Aksamija13,Cahill14}; the former assumes a perfectly smooth interface and only considers the acoustic mismatch between the two materials, while the latter assumes complete randomization of momentum after phonons hit the interface. As most III-V based QCLs are grown by MBE or MOCVD, both well-controlled techniques allowing consistent atomic-level precision, neither AMM nor DMM captures the essence of a III-V SL interface. Figure~\ref{fig:SLsche} shows a schematic of interface roughness in a lattice-matched SL. The jagged dashed boundaries depict transition layers of characteristic thickness $\Delta$ between the two materials.

We introduce a simple model that calculates a more realistic ITBR (a key part in calculating $\kappa_\perp$) by interpolating between the AMM and DMM transmission rates using a specularity parameter $p_\mathrm{spec}$. The model has a single fitting parameter: the effective interface rms roughness $\Delta$. Since the growth environment is well controlled, using one $\Delta$ to describe all the interfaces is justified. We use $\Delta$ to calculate a momentum-dependent specularity parameter
\begin{equation}
p_\mathrm{spec}(\vec{q})=\exp(-4\Delta^2|\vec{q}|^2\cos^2\theta),
\end{equation}
where $|\vec{q}|$ is the magnitude of the phonon wave vector and $\theta$ is the angle between $\vec{q}$ and the normal direction to the interface. Consistent with the twofold impact of interface roughness, $\Delta$ affects the thermal conductivity through two channels. Apart from calculating the ITBR, an effective interface scattering rate $\tau_\mathrm{interface}^{-1}(\vec{q})$ dependent on the same specularity parameter $p_\mathrm{spec}(\vec{q})$ is added to the internal scattering rate to calculate modified $\kappa_\parallel$ (see detailed derivations in \cite{Mei15}). By adjusting only $\Delta$, typically between 1-2 \AA, the calculated thermal conductivity using this model fits a number of different experiments \cite{Yao87,Yu95,Capinski96,Capinski96}.

\subsubsection{$\mathbf{\kappa_\parallel}$ and $\mathbf{\kappa_\perp}$ of a QCL Active Core}
\begin{figure}
	{\includegraphics[width=\columnwidth]{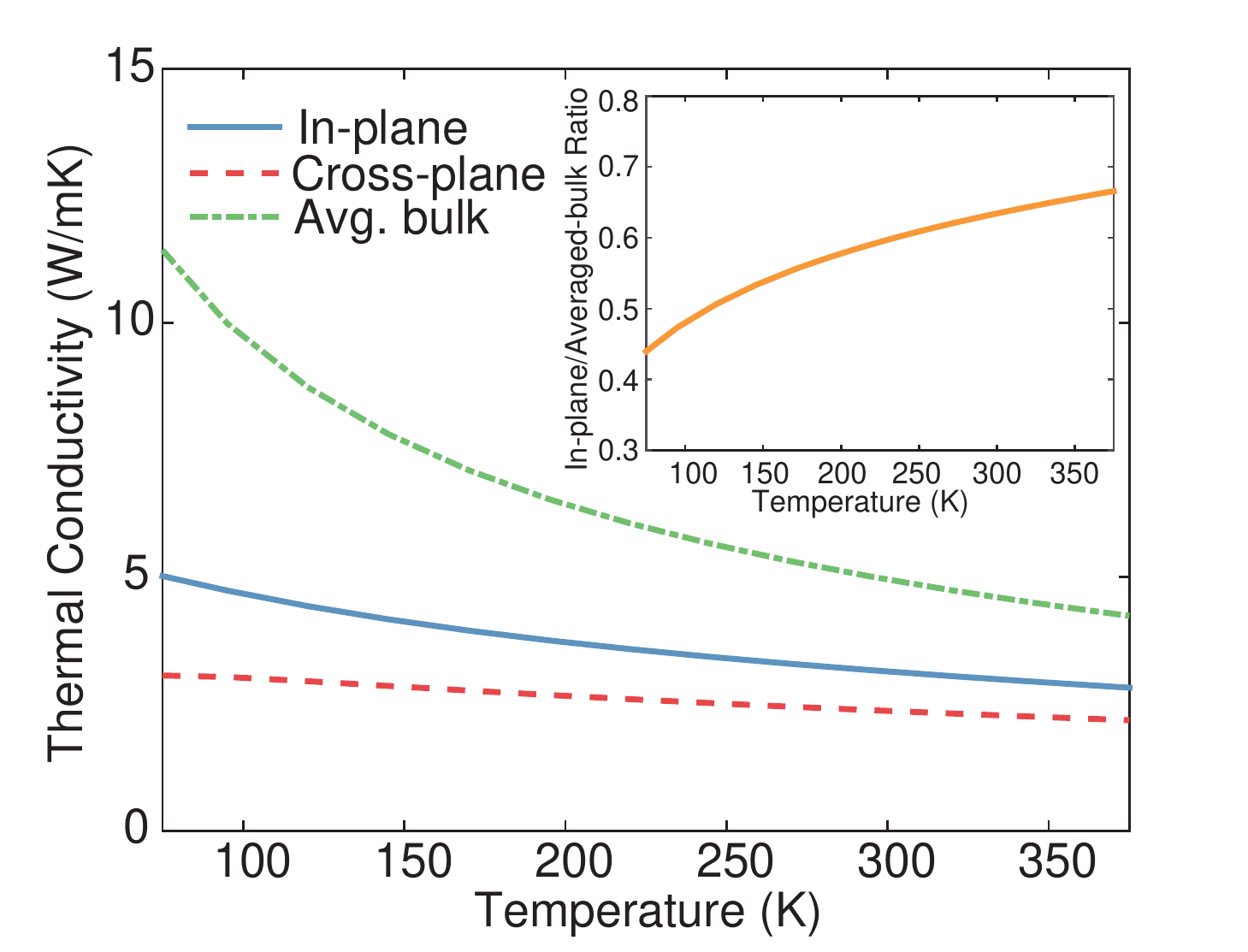}}
	\caption{Thermal conductivity of a typical QCL active region \cite{Lops06} as a function of temperature. A single stage consists of 16 alternating layers of In$_{0.53}$Ga$_{0.47}$As and In$_{0.52}$Al$_{0.48}$As. The solid curve, dashed curve, and dashed-dotted curve show the calculated in-plane, cross-plane, and averaged bulk thermal conductivity, respectively. $\Delta=1\, \AA$ in the calculations. The inset shows the ratio between the calculated in-plane and the averaged bulk thermal conductivities. Figure reproduced from \cite{Mei15}, S. Mei and I. Knezevic, J. Appl. Phys. 118, 175101 (2015), with the permission of AIP Publishing.}\label{fig:SLtc}
\end{figure}

Thermal transport inside the active core of a QCL is usually treated phenomenologically: $\kappa_\parallel$ is typically assumed to be 75\% of the weighted average of the bulk thermal conductivities of the constituent materials, while $\kappa_\perp$ is treated as a fitting parameter (constant for all temperatures) to best fit the experimentally measured temperature profile \cite{Lops06,Lee10}. We calculated the thermal-conductivity tensor of a QCL active core \cite{Lops06} and showed that the typical assumption is not accurate and that the degree of anisotropy is temperature dependent (Fig.~\ref{fig:SLtc}).

The ratio between $\kappa_\parallel$ and the averaged bulk value (inset to Fig.~\ref{fig:SLtc}) varies between 45\% and 70\% over the temperature range of interest. $\kappa_\perp$ has a weak dependence on temperature, in keeping with the common assumption in simplified models; the weak temperature sensitivity means that ITBR dominates cross-plane thermal transport. These results show that it is important to carefully calculate the thermal-conductivity tensor in QCL thermal simulation and we will use this thermal-conductivity model in the device-level simulation.

%%%%%%%%%%%%%%%%%%%%%%%%%%%%%%%%%%%%%%%%%%%%%%%%%%
\subsubsection{Other Materials}
%%%%%%%%%%%%%%%%%%%%%%%%%%%%%%%%%%%%%%%%%%%%%%%%%%
The active core is not the only region we need to model in a device-level thermal simulation. Figure~\ref{fig:QCLsche} shows a typical schematic (not to scale) of a QCL device in thermal simulation with a substrate-side mounting configuration \cite{Page01}. The active core (in this case, consisting of 36 stages and 1.6 $\mu$m thick) with width $W_\mathrm{act}$ is embedded between two cladding layers (4.5-$\mu$m-thick GaAs). The waveguide is supported by a substrate (GaAs) with thickness $D_\mathrm{sub}$. An insulation layer (Si$_3$N$_4$) with thickness $D_\mathrm{ins}$ is deposited around the waveguide and then etched away from the top to make the contact. Finally, a contact layer (Au) with thickness $D_\mathrm{cont}$ and a thin layer of solder ($D_\mathrm{sold}$) are deposited on top. There is no heat generation in the regions other than the active core. Further, these layers are typically thick enough to be treated as bulk materials. Bulk-substrate (GaAs or InP) thermal conductivities are readily obtained for III-V materials from experiment, as well as from relatively simple theoretical models \cite{Mei15,Lee10,Lops06,Spagnolo08} (Table~\ref{table:bulktc}).

\begin{table}
\caption{Thermal conductivity as a function of temperature for materials in a QCL structure.}\label{table:bulktc}
	\tabcolsep25pt
	\begin{tabular}{@{}ll@{}}
		Materials & Thermal conductivity (W/mK) \\
		\toprule
		Au & $337-600\times10^{-4}T$ \\
		Si$_3$N$_4$& $30-1.4\times 10^{-2}T$ \\
		In solder & $93.9-6.96\times 10^{-2}T+9.86\times10^{-5}T^2$ \\
	\end{tabular}
\end{table}

%%%%%%%%%%%%%%%%%%%%%%%%%%%%
\subsection{Device-level Electrothermal Simulation}\label{sec:device}

\subsubsection{Device Schematic}
%%%%%%%%%%%%%%%%%%%%%%%%%%%

\begin{figure}
	{\includegraphics[width=\columnwidth]{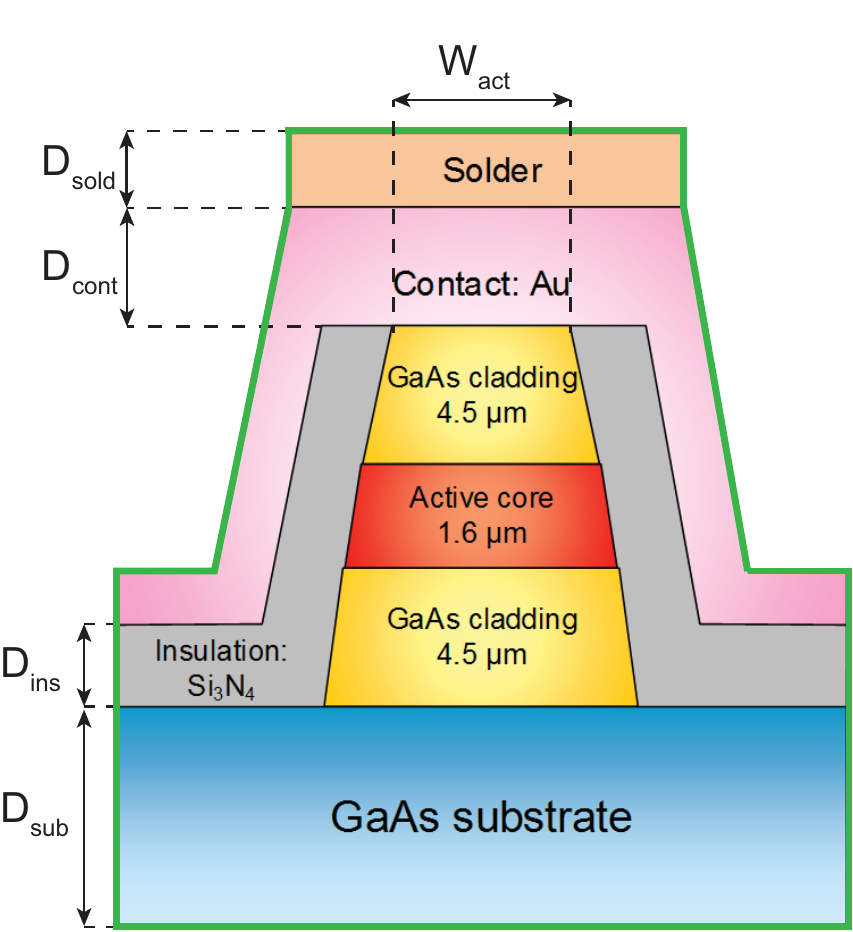}}
	\caption{Schematic of a typical GaAs-based mid-IR QCL structure with a substrate (not to scale).}\label{fig:QCLsche}
\end{figure}

The length of a QCL device is much greater than its width, therefore we can assume the length is infinite and carry out a 2D thermal simulation. The schematic of the simulation domain (not to scale) is shown in Fig.~\ref{fig:QCLsche}. The boundary of the simulation region is highlighted in green, and certain boundary conditions (heat sink at fixed temperature, convective boundary condition, or adiabatic boundary condition) can be applied (independently) to each boundary. Typically, the bottom boundary of the device is connected to a heat sink while other boundaries have the convective boundary condition at the environment temperature (single-device case) or the adiabatic boundary condition (QCL-array case).
Typical values for the layers thickness are $W_\mathrm{act}=15~\mu$m, $D_\mathrm{sub}=50~\mu$m, $D_\mathrm{ins}=0.3~\mu$m, $D_\mathrm{cont}=3~\mu$m, $D_\mathrm{sold}=1.5~\mu$m.

\begin{figure}
	{\includegraphics[width=\columnwidth]{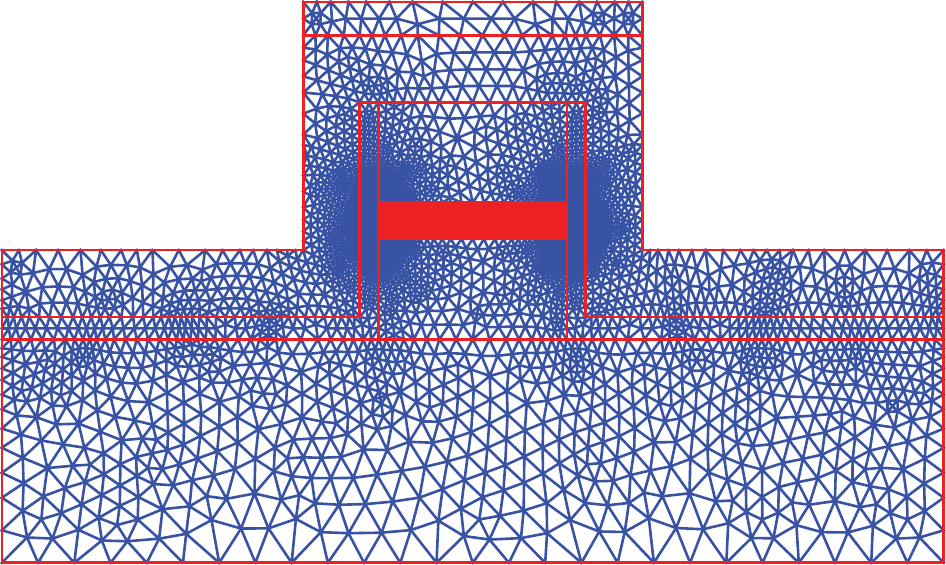}}
	\caption{A typical nonuniform finite-element mesh of the simulated GaAs-based mid-IR QCL structure.}\label{fig:QCLmesh}
\end{figure}

We use the finite-element method to solve for the temperature distribution. The whole device is divided into different regions according to their materials properties. Each stage of the active region is treated as a single unit with the heat-generation rate tabulated in the device table in order to capture the nonuniform behavior among stages. The active core is very small, but is also the only region with heat generation, small thermal conductivity, and spatial nonuniformity. To capture the behavior of the active region while saving computational time, we use a nonuniform mesh in the finite-element solver to emphasize the active core region. Figure~\ref{fig:QCLmesh} shows a mesh generated in the simulation.

%%%%%%%%%%%%%%%%%%%%%%%%%%%%%%
\subsubsection{Simulation Algorithm}
%%%%%%%%%%%%%%%%%%%%%%%%%%%%%%

It is known that among all the stages in the active core, the temperature $T_i$ and the electric field $F_i$ ($i$ represents the stage index) are not constant \cite{Wacker02,Lops06}, but we have no \textit{a priori} knowledge of how they depend on the stage index. However, we know that the charge--current continuity equation must hold, and in the steady state $\nabla\cdot\mathbf{J}=0$; this implies that the current density $J$ must be uniform, as the current flow is essentially in one dimension, along $z$. This insight is key to bridging the single-stage and device-level simulations.

From Sec.~\ref{sec:EMC}, we can obtain the heat-generation rate $Q$ inside the active core by running the single-stage EMC simulation. Each single-stage EMC is carried out at a specific electric field $F$ and temperature $T$ and outputs both the current density $J(F,T)$ and the heat-generation rate $Q(F,T)$. By sweeping $F$ and $T$ in range of interest, we obtain a table connecting different field and temperature $(F,T)$ to proper current density and heat-generation rate $(J,Q)$ [$(F,T)\rightarrow(J,Q)$]. However, from the discussion above, the input in the thermal simulation needs to be the constant parameter $J$. Therefore, we ``flip'' the recorded $(F,T)\rightarrow(J,Q)$ table to a so-called device table $(J,T)\rightarrow(Q,F)$, suitable for coupled simulation~\cite{Shi16_FdP}.

\begin{figure}
	{\includegraphics[width=5 cm]{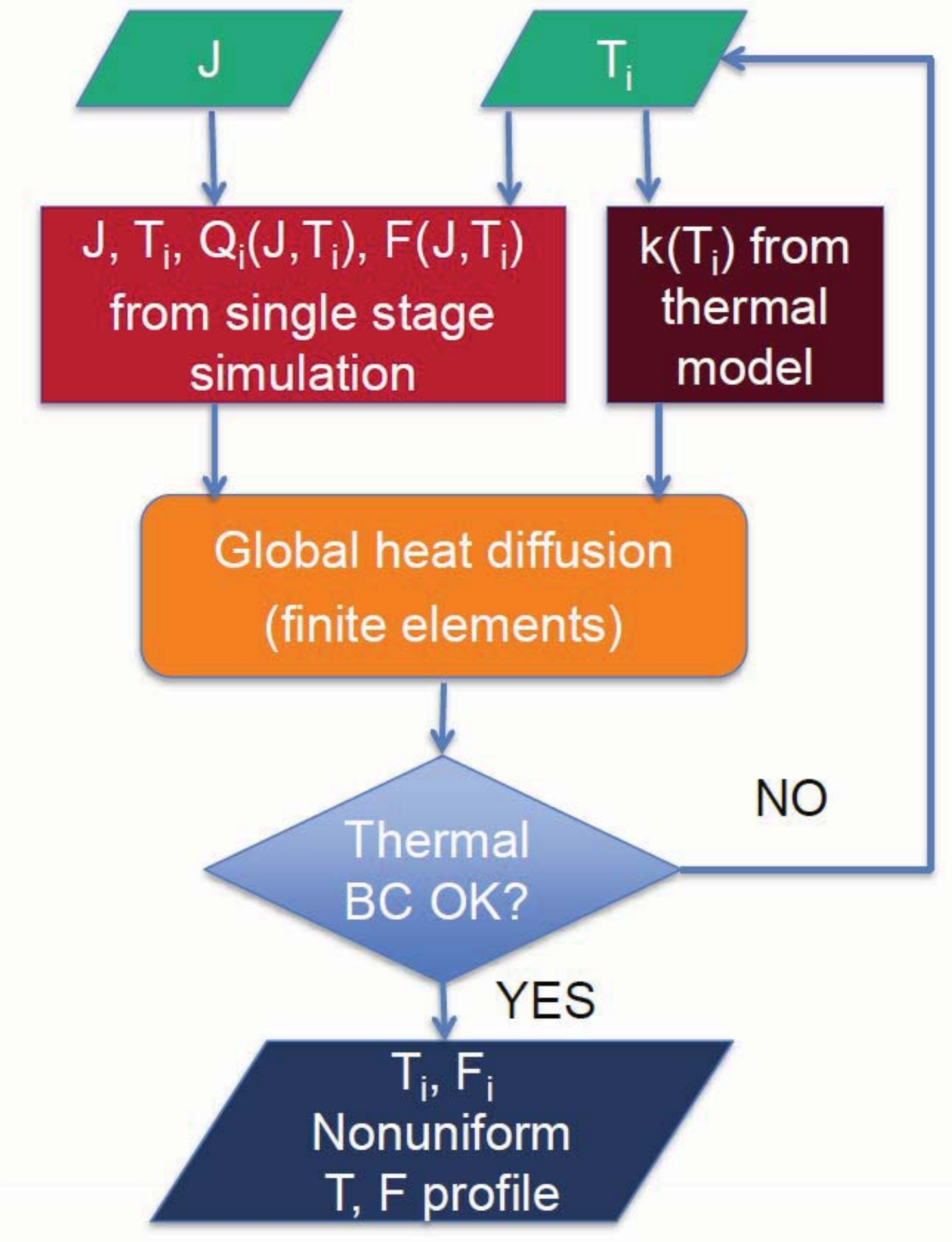}}
	\caption{Flowchart of the device-level thermal simulation.  We start by assuming a certain current density $J$ and temperature profile $T_i$ across the whole device. Based on the tabulated information from the single-stage simulation and assumed $(J,T_i)$, we get stage-by-stage profile for the electric field $F_i$ and the heat-generation rate $Q_i$ profiles. An accurate temperature-dependent thermal conductivity model, which includes the
boundary resistances of layers, and the temperature profile guess are used as input to the heat diffusion equation, which is then iteratively solved (with updated temperature profile in each step) until the thermal boundary conditions are satisfied.}\label{fig:Thermalflchart}
\end{figure}

Figure~\ref{fig:Thermalflchart} depicts the flowchart of the device-level electrothermal simulation \cite{Shi16_FdP}. Before the simulation, we obtain the device table [$(J,T)\rightarrow(Q,F)$], as discussed above. We also have to calculate the thermal conductivities ($\kappa_\parallel$ and $\kappa_\perp$) of the active region as a function of temperature and tabulate them, based on the model described in Sec.~\ref{sec:IIIVSL}. We also need the bulk thermal conductivity of other materials in the device (cladding layer, substrate, insulation, contact, and solder) as a function of temperature. These material properties are standard and already well characterized.

Each device-level thermal simulation is carried out in a certain environment (i.e., for a given set of boundary conditions) and with a certain current density $J$. At the beginning of the simulation, an initial temperature profile is assigned. With the input from the device table and the thermal conductivity data in each region, we use a finite-element method to iteratively solve the heat diffusion equation until convergence. At the end of the simulation, we obtain a thermal map of the whole device. Further, from the temperature $T_i$ in each stage and the injected current density $J$, we obtain the nonuniform electric field distribution $F_i$. With the electric field in each stage and given the stage thickness, we can accurately calculate the voltage drop across the device and obtain the current--voltage characteristic. By changing the mounting configuration ($W_\mathrm{act}$, $D_\mathrm{sub}$, $D_\mathrm{ins}$, $D_\mathrm{cont}$, $D_\mathrm{sold}$) or the boundary conditions, the temperature profile can be changed.

%*****************************************************************************************
% Section 4, Example on the 9 um device

%%%%%%%%%%%%%%%%%%%%%%%%%%%%%%%%%%%%%%%%%%%%%%
%%%%%%%%%%%%%%%%%%%%%%%%%%%%%%%%%%%%%%%%%%%%%%
\section{Device-level Electrothermal Simulation: An Example}\label{sec:example}
%%%%%%%%%%%%%%%%%%%%%%%%%%%%%%%%%%%%%%%%%%%%%%
%%%%%%%%%%%%%%%%%%%%%%%%%%%%%%%%%%%%%%%%%%%%%%

In this section, we present detailed simulation results of a 9-$\mu$m GaAs/Al$_{0.45}$Ga$_{0.55}$As mid-IR QCL \cite{Page01} based on a conventional three-well active region design. The chosen structure has 36 repetitions of the single stage; each stage has 16 layers. Starting from the injection barrier, the layer thicknesses in one stage (in $\AA$) are \textbf{46}/19/\textbf{11}/54/\textbf{11}/48/\textbf{28}/34/\textbf{17}/30/\textbf{\underline{18}}/\underline{28}/\textbf{\underline{20}}/\underline{30}/\textbf{26}/30.  Here, the barriers  (Al$_{0.45}$Ga$_{0.55}$As)   are in bold while the wells (GaAs)   are in normal font; the underlined layers are doped to a sheet density of $n_{\mathrm{Si}}=3.8\times10^{11}\,\mathrm{cm}^{-2}$. The results at 77~K are shown here.

\subsection{Electronic Simulation Results}\label{sec:elresult}
\subsubsection{Band Structure}

\begin{figure}
	{\includegraphics[width=\columnwidth]{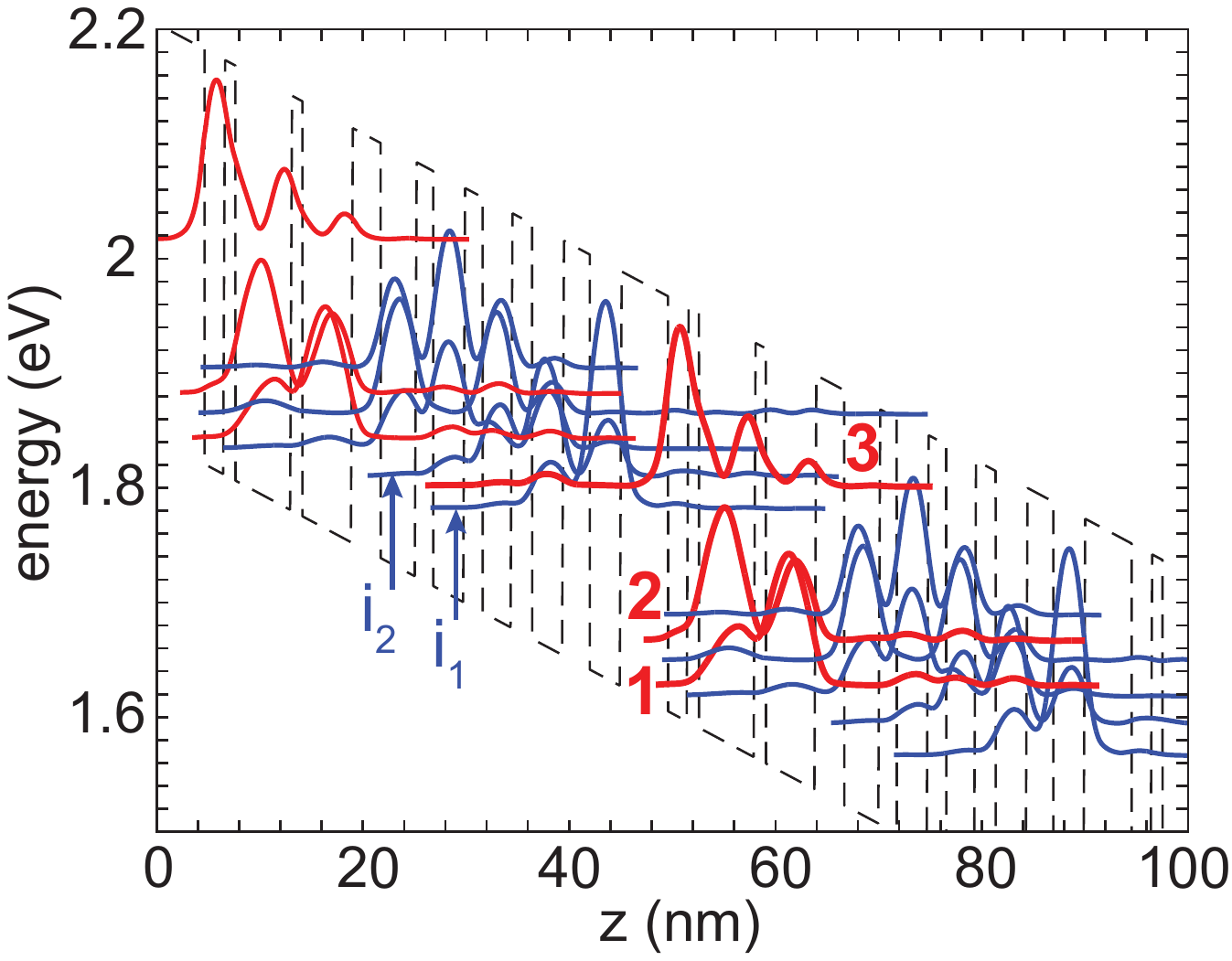}}
	\caption{Energy levels and wavefunction moduli squared of $\Gamma$-valley subbands in two adjacent stages of the simulated GaAs/AlGaAs-based structure. The bold red curves denote the active region states (1, 2, and 3 represent the ground state and the lower and upper lasing levels, respectively). The blue curves represent injector states, with $i_1$ and $i_2$ denoting the lowest two. Figure reproduced from \cite{Shi14}, Y. B. Shi and I. Knezevic, J. Appl. Phys. 116, 123105 (2014), with the permission of AIP Publishing.}\label{fig:pageband}
\end{figure}

Figure~\ref{fig:pageband} shows the electronic states of the chosen structure under the design operating field of 48 kV/cm, calculated from the coupled $\mathbf{k\cdot p}$--Poisson solver (see Sec.~\ref{sec:kp}). The active-region states of the central stage are represented in bold red curves; 1, 2, and 3 are the ground state and the lower and upper lasing levels, respectively. Injector states are labeled $i_1$ and $i_2$. Other blue states together form the miniband. (When the electron density in the QCL is high, the electronic bands have to be calculated self-consistently with EMC.)

\subsubsection{$J-F$ Curve}\label{sec:JF}
The current density $J$ vs field $F$ curve, one of the key QCL characteristics at a given temperature, is intuitive to obtain in EMC. After calculating the electronic band structure at a certain field $F$, the wavefunctions, energy levels, and effective masses of each subband and each stage are fed into the EMC solver. In the EMC simulation, we include all the scattering mechanisms described in Sec.~\ref{sec:EMC}. Since we employ periodic boundary conditions, the current density $J$ can be extracted from how many electrons cross the stage boundaries in a certain amount of time in the steady state. The net flow $n_\mathrm{net}$ of electrons is calculated by subtracting the flow between the central stage and the previous stage ($n_\mathrm{backward}$) from the flow between the central stage and the next stage ($n_\mathrm{forward}$) in each time step. The current density is then calculated as
\begin{equation}\label{eq:curden}
J=\frac{en_\mathrm{net}}{A_\mathrm{eff}\delta t}=\frac{e(n_\mathrm{forward}-n_\mathrm{backward})}{A_\mathrm{eff}\delta t}\, ,
\end{equation}
where $\delta t$ is the time interval during which the flow is recorded. $A_\mathrm{eff}$ is the effective in-plane area of the simulated device. Since doping is the main source of electrons, the area is calculated as
\begin{equation}\label{eq:Aeff}
A_\mathrm{eff}=\frac{N_\mathrm{ele}}{N_s}\, ,
\end{equation}
where $N_\mathrm{ele}$ is the number of simulated electrons and $N_s$ is the sheet doping density (in cm$^{-2}$) in the fabricated device. In the current simulation, $N_\mathrm{ele}=50,000$ and $N_s=3.8\times 10^{11}~\mathrm{cm}^{-2}$.

Due to the stochastic nature of EMC, we need to average the current density over multiple time steps. In practice, one can record the net cumulative number of electrons per unit area that leave a stage over time and obtain a linear fit to this quantity in the steady state; the slope yields the steady-state current density.

\begin{figure}
	{\includegraphics[width=\columnwidth]{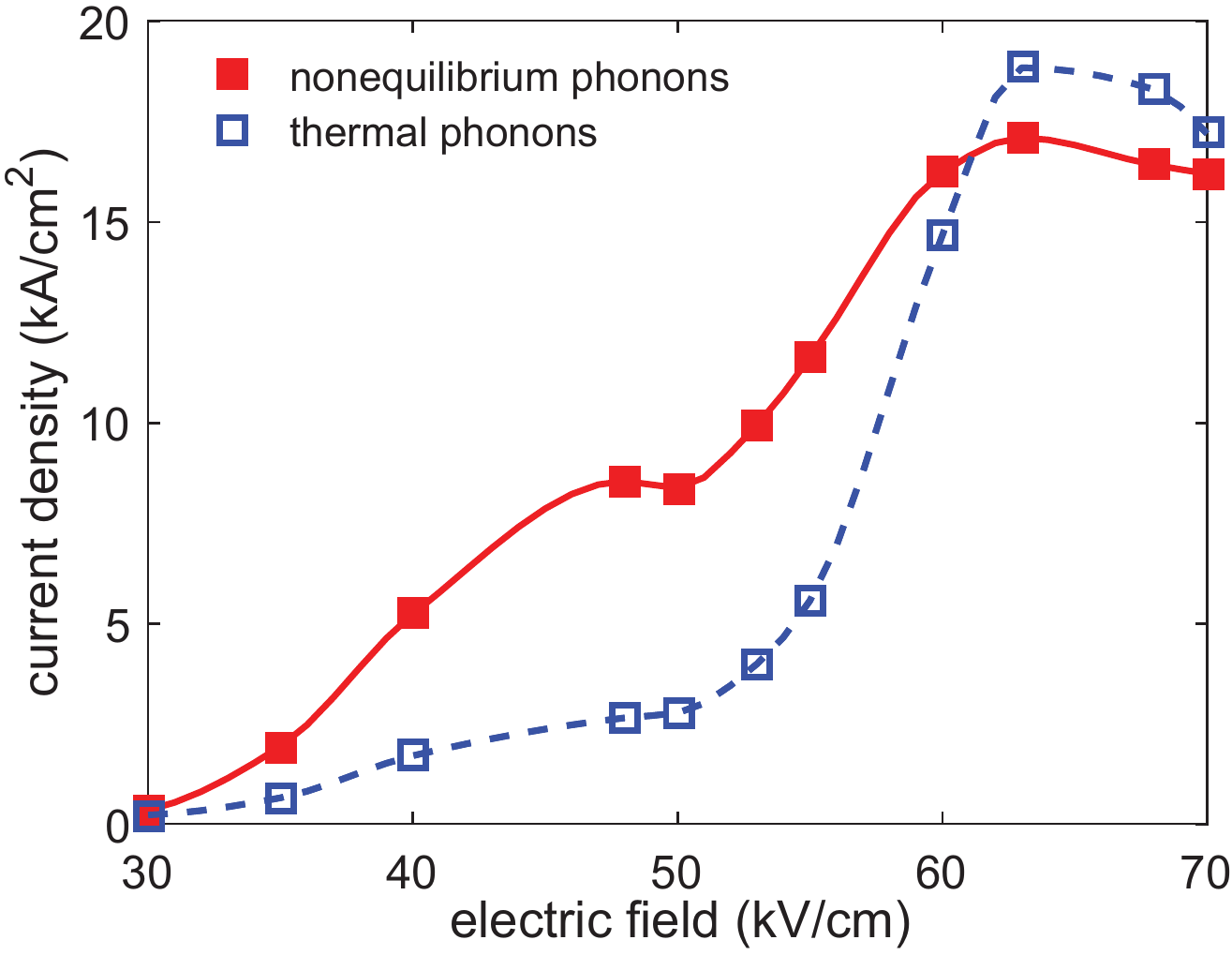}}
	\caption{The current density vs. electric field ($J-F$) curve of the simulated device with   (filled squares)  and without  (empty squares)   the nonequilibrium phonon effect at 77~K.   The inclusion of nonequilibrium phonons  considerably increases the current density at a given field up to 60 kV/cm. }\label{fig:JF}
\end{figure}

From each individual simulation, we extract the current density at a given electric field and temperature. To obtain the $J-F$ curve at that temperature, we sweep the electric field.  To demonstrate the importance of including nonequilibrium phonons effects, we carry out the simulation with thermal phonons alone and with both thermal and excess nonequilibrium phonons. Figure~\ref{fig:JF} is the $J-F$ curve for the simulated structure with (filled squares)  and without   (empty squares) nonequilibrium phonons at 77~K. It can be seen that the current density at a given field considerably increases when nonequilibrium phonons are included and the trend holds up to 60 kV/cm. This difference is prominent at low temperatures (< 200~K) and goes away at RT \cite{Shi14} .

\subsubsection{Modal Gain ($G_m$) and Threshold}\label{sec:modalgain}
We calculate the modal gain as \cite{Mircetic05}
\begin{equation}\label{eq:modalgain}
G_m=\frac{4\pi e^2\left<z_{32}\right>^2\Gamma_w\Delta n}{2\varepsilon_0\underline{n}\gamma_{32}L_p\lambda}\, ,
\end{equation}
where $\varepsilon_0$ is the permittivity of free space. Some constants are obtained from experiment: waveguide confinement factor $\Gamma_w=0.31$, stage length $L_p=45$ nm, optical-mode refractive index $\underline{n}=3.21$, and full width at half maximum $\gamma_{32}(T_L)\approx8.68~\mathrm{meV}+0.045~\mathrm{meV/K}\times T_L$ \cite{Shi14,Page01}. The dipole matrix element between the upper and lower lasing levels ($\left<z_{32}\right>=1.7$ nm) and the emission wave length ($\lambda=9~\mu$m) are also estimated in experiment \cite{Page01}, but we calculate these two terms directly. The dipole matrix element is calculated as
\begin{equation}\label{eq:dipole}
\left<z_{32}\right>=\int_{0}^{d} z\varphi_3^*(z)\varphi_2(z)dz\, .
\end{equation}
The value is slightly different at different fields, as the band structure changes. At 48 kV/cm, the calculated matrix element is $\left<z_{32}\right>=1.997$ nm.
Similarly, the wave length of emitted photon also changes at different fields. One can calculate the value from the energy difference between the upper and lower lasing levels. The calculated wave length at 48 kV/cm is 8.964 $\mu$m.
$\Delta n=n_\mathrm{upper}-n_\mathrm{lower}$ is the population inversion obtained from EMC. Again, due to the randomness of EMC, the population inversion needs to be averaged over a period of time after the steady-state has been reached.

\begin{figure}
	{\includegraphics[width=\columnwidth]{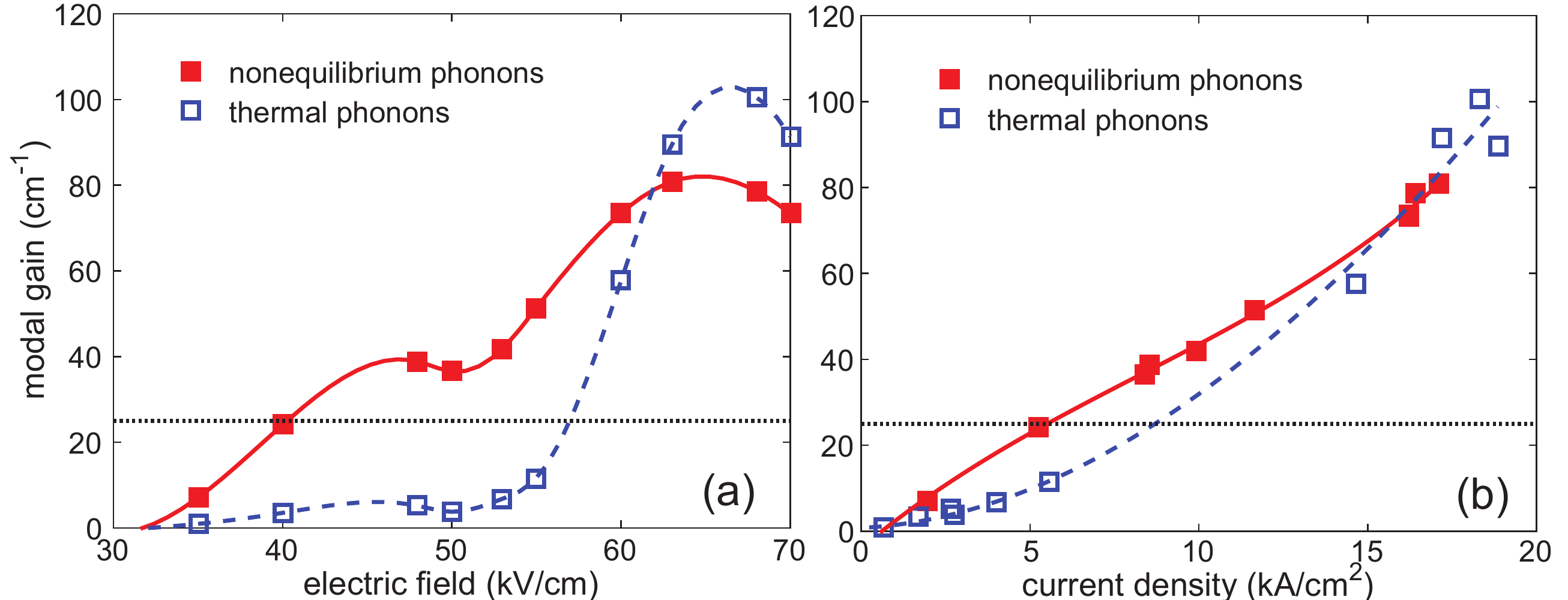}}
	\caption{The modal gain ($G_m$) of the simulated device with nonequilibrium (filled squares) and thermal (empty squares) phonons as a function of (a) applied electric field and (b) current density $J$ at 77~K. Horizontal dotted line shows the total estimated loss of the device.}\label{fig:gm}
\end{figure}

Figure~\ref{fig:gm} shows the modal gain of the device with nonequilibrium (filled squares) and thermal (empty squares) phonons as a function of (a) electric field and (b) current density at 77~K. Horizontal dotted line indicates the total estimated loss in the device, which is used to help find the threshold current density, $J_\mathrm{th}$. Lasing threshold is achieved when the modal gain $G_m$ equals the total loss $\alpha_\mathrm{tot}$. We consider two sources of loss, mirror ($\alpha_\mathrm{m}$) and waveguide ($\alpha_\mathrm{w}$), so the total loss is $\alpha_\mathrm{tot}=\alpha_\mathrm{m}+\alpha_\mathrm{w}$. The intercepts between the total loss line and the $G_m$ vs $F$ [Fig.~\ref{fig:gm}(a)] and $G_m$ vs $J$ [Fig.~\ref{fig:gm}(b)] curves give the threshold field $F_\mathrm{th}$ and threshold current density $J_\mathrm{th}$, respectively. Like the current density, the modal gain of the device is also considerably higher when nonequilibrium phonons are considered, which leads to a lower $F_\mathrm{th}$ and a lower $J_\mathrm{th}$. The reason for the increased current density and modal gain with nonequilibrium phonons can be attributed to the enhanced injection selectivity and efficiency \cite{Shi14}.

\subsubsection{Heat-generation Rate}\label{sec:heatgen}

\begin{figure}
	{\includegraphics[width=\columnwidth]{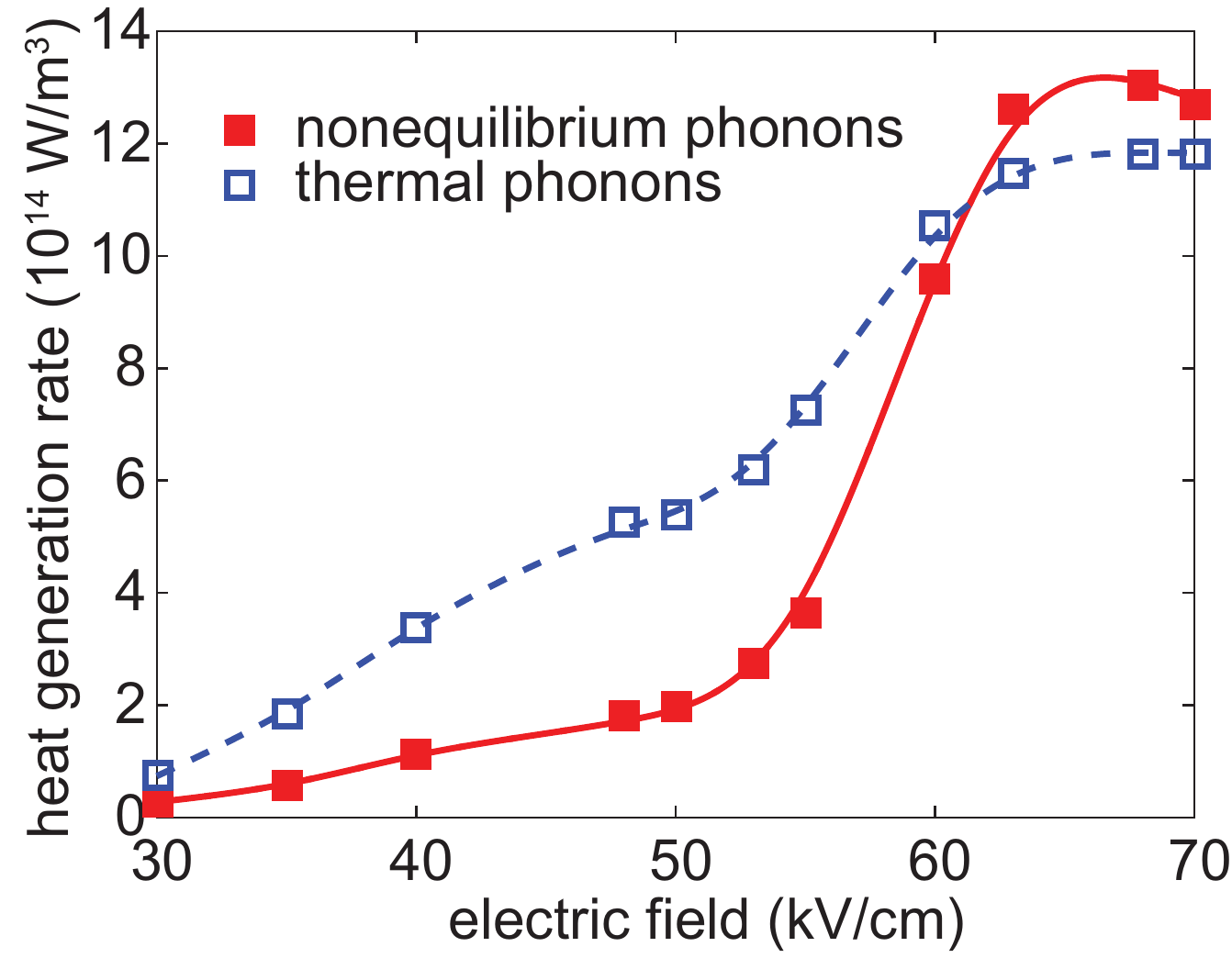}}
	\caption{The heat-generation rate of the simulated device as a function of electric field $F$ at 77~K with   (filled squares)   and without   (empty squares)   nonequilibrium phonons.}\label{fig:hgen}
\end{figure}

The way to obtain the heat-generation rate $Q$ is similar to how we get the current density $J$. We record the cumulative net energy emission as a function of time and fit a straight line to the region where the simulation has reached a steady state. The slope of the line is used in place of $\frac{\sum(\hbar\omega_\mathrm{ems}-\hbar\omega_\mathrm{abs})}{t_\mathrm{sim}}$. Figure~\ref{fig:hgen} shows the heat-generation rate as a function of electric field at 77~K. The   filled squares and the empty squares depict the situation with and without nonequilibrium phonons, respectively.

%%%%%%%%%%%%%%%%%%%%%%%%%%%%%%%%%%%%%%%%%%%%%%%%%%
%%%%%%%%%%%%%%%%%%%%%%%%%%%%%%%%%%%%%%%%%%%%%%%%%%
\subsection{Representative Electrothermal Simulation Results}\label{sec:thresult}
%%%%%%%%%%%%%%%%%%%%%%%%%%%%%%%%%%%%%%%%%%%%%%%%%%
%%%%%%%%%%%%%%%%%%%%%%%%%%%%%%%%%%%%%%%%%%%%%%%%%%

This section serves to illustrate how the described simulation is implemented in practice, and what type of information it provides at the single-stage and device levels.

\begin{figure}
	{\includegraphics[width=\columnwidth]{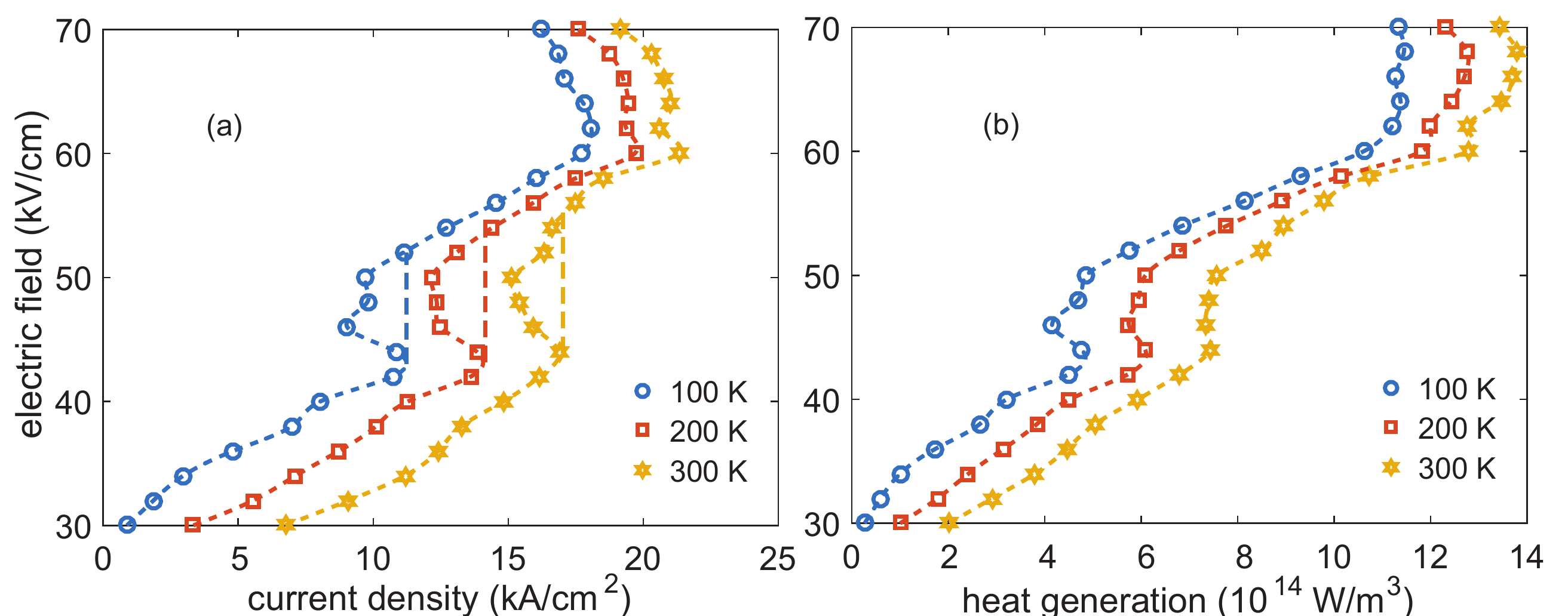}}
	\caption{The field vs. current density (a) and heat-generation rate vs. current density (b) characteristics for the simulated device at 100, 200, and 300 K, as obtained from single-stage simulation with nonequilibrium phonons.}\label{fig:SingleStageJF}
\end{figure}

First, the single-stage coupled simulation has to be performed at different temperatures, as in Fig.~\ref{fig:SingleStageJF}(a). We note the the calculated $J-F$ curves show a negative-differential-conductance region, which is typical for calculations, but generally not observed in experiment. Instead, a flat $J-F$ dependence is typically recorded \cite{Dupont10}. At every temperature and field, we also record the heat-generation rate, as depicted in Fig.~\ref{fig:SingleStageJF}(b).

\begin{figure}
	{\includegraphics[width=\columnwidth]{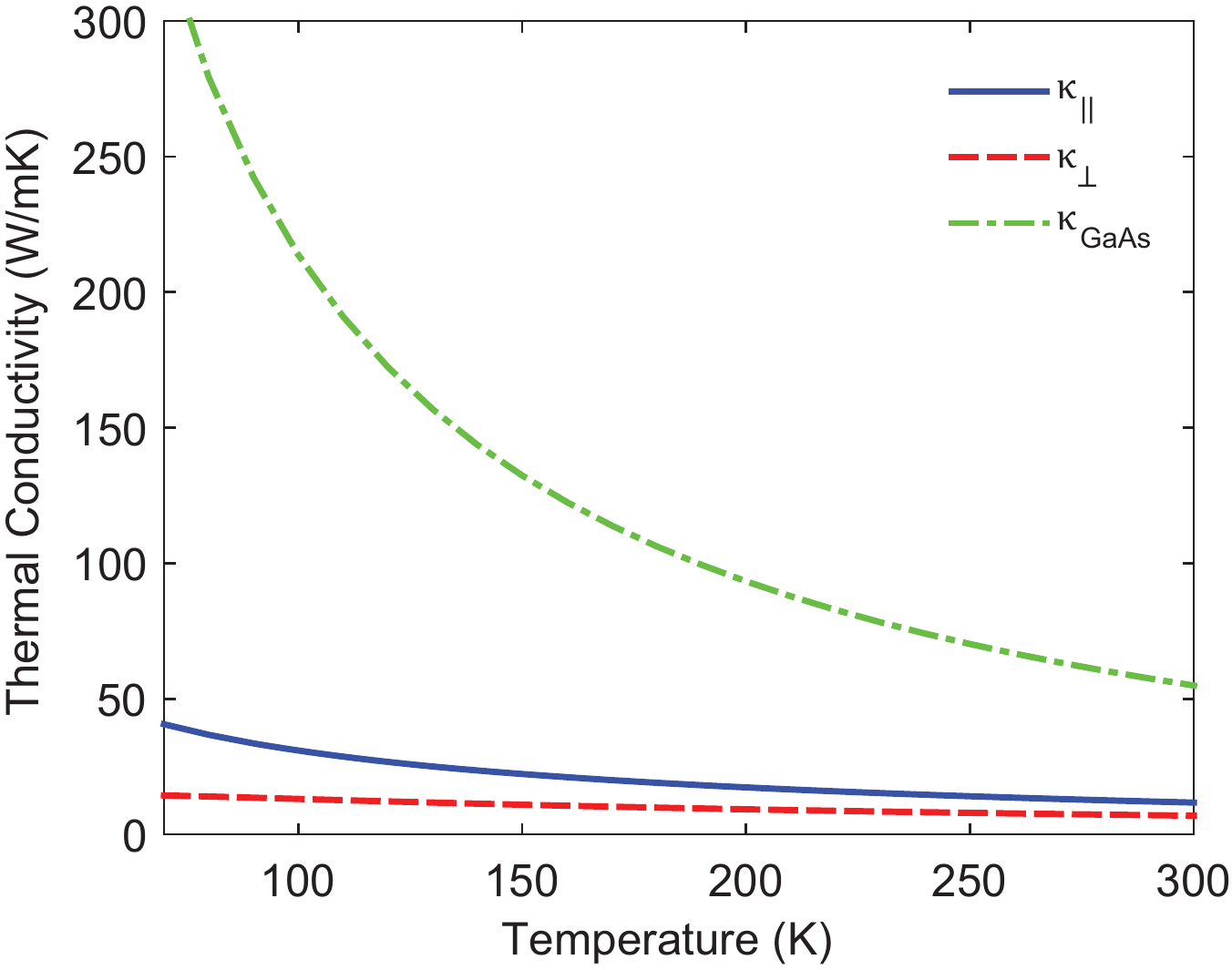}}
	\caption{Calculated in-plane ($\kappa_\parallel$; solid line) and cross-plane ($\kappa_\perp$; dashed line) thermal conductivities of the active core, along with the bulk thermal conductivity of the GaAs substrate (dash-dotted line). The effective rms roughness $\Delta$ is taken to be $5\, \AA.$}\label{fig:Pagetc}
\end{figure}

Second, the thermal model for the whole structure is developed. Considering that growth techniques improve over time, structures grown  around the same time should have similar properties. Since the device studied here was built in 2001 \cite{Page01}, we assume the active core should have similar effective rms roughness $\Delta$ to other lattice-matched GaAs/AlAs SLs built around the same time \cite{Capinski96,Capinski99}. From our previous simulation work on fitting the SL thermal conductivities \cite{Mei15}, we choose an effective rms roughness $\Delta=5\, \AA$ in this calculation. Figure~\ref{fig:Pagetc} shows the calculated thermal conductivities $\kappa_\parallel$ (solid line) and $\kappa_\perp$ (dashed line), along with the calculated bulk thermal conductivity (dash-dotted line) for the substrate GaAs.

\begin{figure}
{\includegraphics[width=8cm]{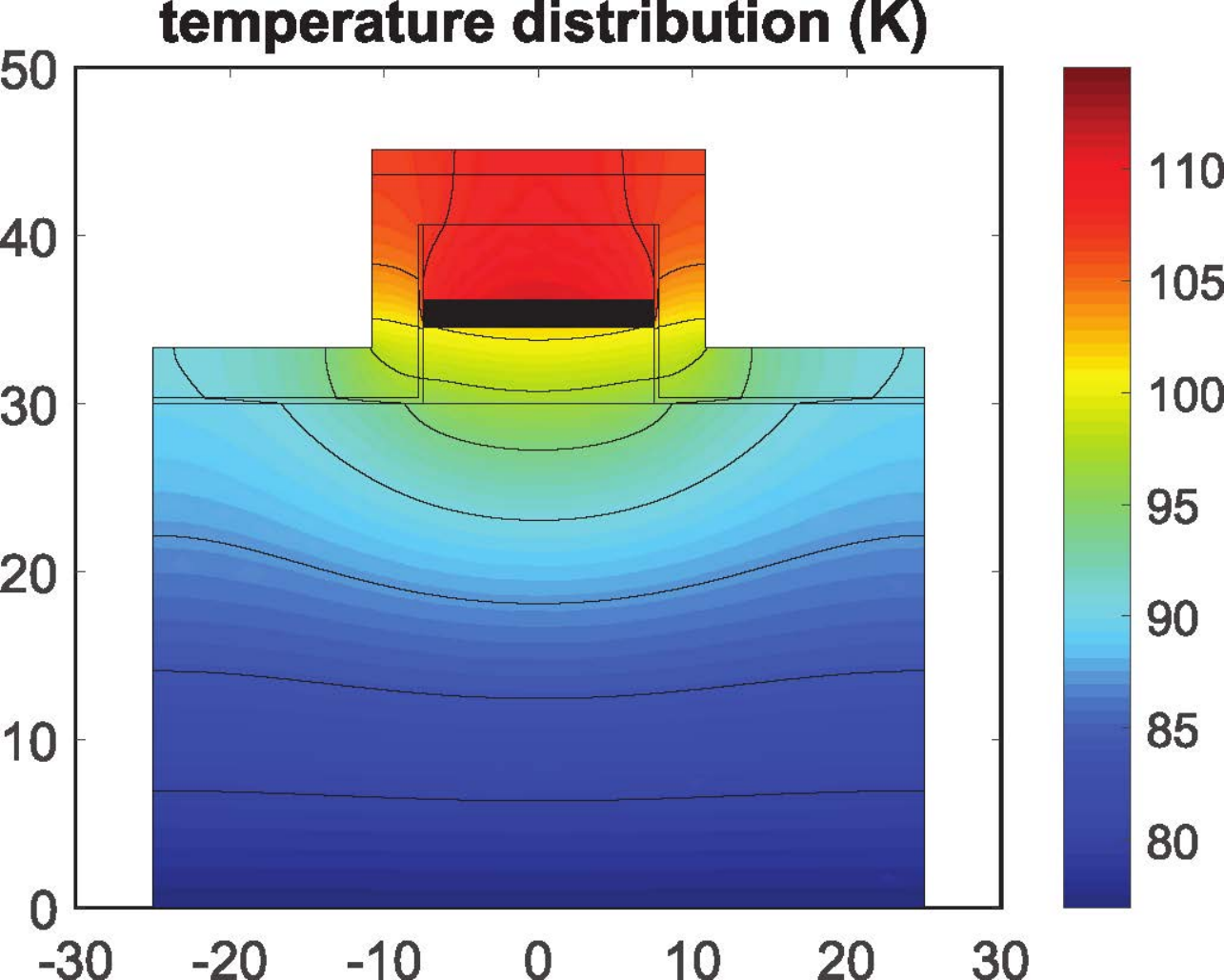}}
	\caption{A typical temperature profile across the structure. At the bottom of the device is a heat sink held at 77~K, while adiabatic boundary conditions are applied elsewhere. The current density is $6\,\mathrm{kA/cm^2} $ and the duty cycle is 100\%.}\label{fig:TempProfileSchematic}
\end{figure}

\begin{figure}
{\includegraphics[width=\columnwidth]{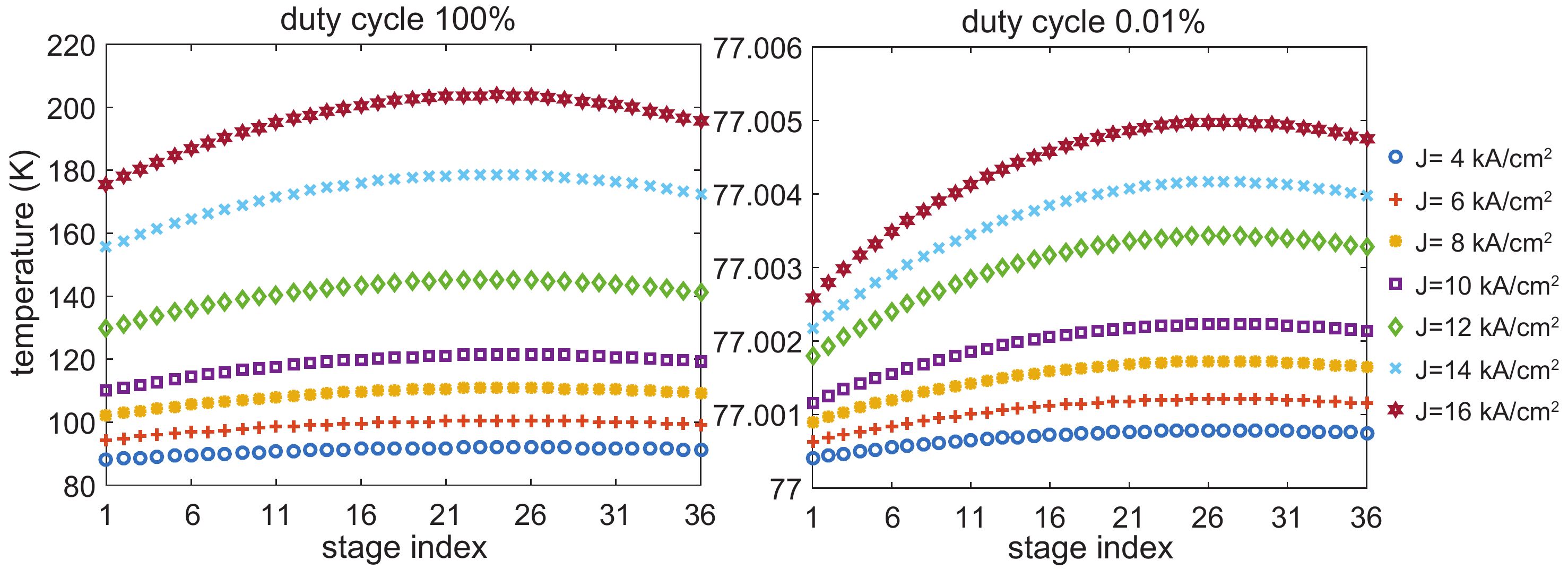}}
	\caption{Temperature profile inside the active region at 100\% duty cycle (left) and 0.01\% duty cycle (right) for the QCL of Page \textit{et al.} \cite{Page01}.}\label{fig:TempProfileDutyCycle}
\end{figure}

\begin{figure}	
{\includegraphics[width=\columnwidth]{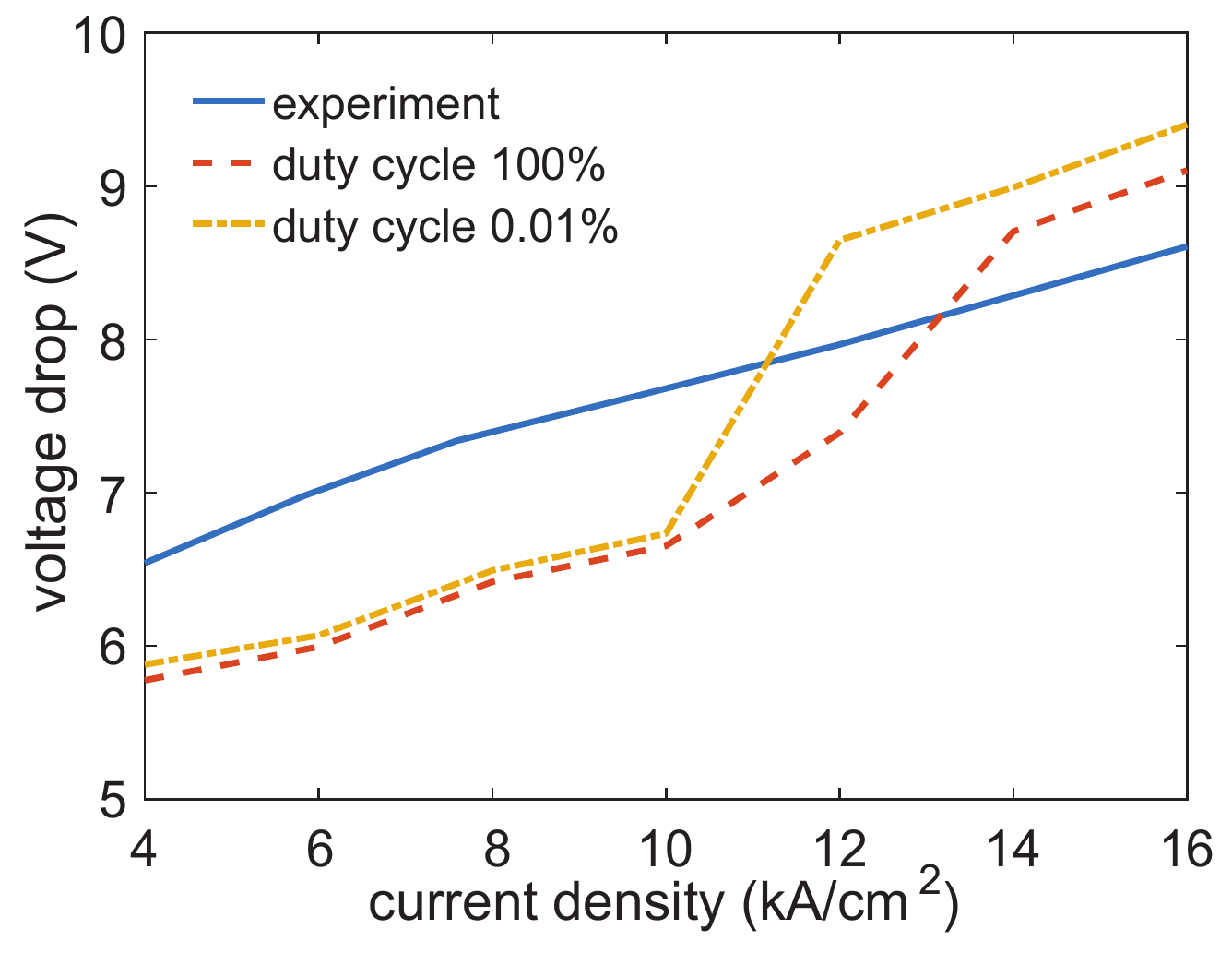}}
	\caption{The current density vs. voltage drop for the simulated device in experiment (solid curve) and as calculated at 100\% (dashed curve) and 0.01\% (dot-dashed curve) duty cycles. The bottom of the device is placed on a heat sink held at 77~K while adiabatic boundary conditions are assumed on the rest of the boundaries (see Sec.~\ref{sec:device}).}\label{fig:JV}
\end{figure}

The structure we considered operated in pulsed mode at 77~K. Depending on the duty cycle, the temperature distribution in the device can differ considerably. Figure~\ref{fig:TempProfileSchematic} depicts a typical temperature profile across the device, while Fig.~\ref{fig:TempProfileDutyCycle} depicts the profile across the active core alone at duty cycles of 100\% (essentially continuous wave lasing, if the device achieved it) and 0.01\% (as in experiment \cite{Page01}). Clearly, CW operation would results in dramatic heating of the active region. Finally, Fig.~\ref{fig:JV} shows the $J-V$ curve of the entire simulated device at 77~K with a duty cycles of 0.01\%, 100\%, and as observed in experiment \cite{Page01}.

\section{Conclusion}
 We overviewed electronic and thermal transport simulation of QCLs, as well as recent efforts in device-level electrothermal modeling of these structures, which is appropriate for transport below threshold, where the effects of the optical field are negligible. We specifically focused on mid-IR QCLs in which electronic transport is largely incoherent and can be captured by the ensemble Monte Carlo technique. The future of QCL modeling, especially for near-RT CW operation, will likely include improvements on several fronts: 1) further development of computationally efficient yet rigorous quantum-transport techniques for electronic transport, to fully account for coherent transport features that are important in short-wavelength mid-IR devices; 2) a better understanding and better numerical models for describing the role of electron--electron interaction, impurities, and interface roughness on device characteristics; 3) holistic modeling approaches in which electrons, phonons, and photons are simultaneously and self-consistently captured within a single simulation. The goal of QCL simulation should be nothing less than excellent predictive value of device operation across a range of temperatures and biasing conditions, along with unprecedented insight into the fine details of exciting nonequilibrium physics that underscores the operatiuon of these devices.

\section{Acknowledgement}
The authors gratefully acknowledge support by the U.S. Department of Energy, Basic Energy Sciences, Division of Materials Sciences and Engineering, Physical Behavior of Materials Program, Award No. DE-SC0008712. The work was performed using the resources of the UW-Madison Center for High Throughput Computing (CHTC).

%*****************************************************************************************
% the bibligraphy

%\bibliographystyle{apsrev}
%\bibliography{mybib}

%\putbib
%%%%%%%%%%%%%%%%%%%%%%%%%%%%%%%%%%%%%%%%%%%%%%%%%%%%%%%%%%%%%%%%%%%%%%%%%%%%%%%%%%%%%%%%%%%%%%%%%%%%%%%%%%%%%%%%%%%%
\end{document}